\documentclass[useAMS,usenatbib,usegraphicx]{mn2e}
\usepackage{pdflscape}
\usepackage{graphicx}
\usepackage{longtable}
\usepackage{lscape}
\usepackage{amssymb}
\usepackage{endnotes} 
\usepackage{footnote}
\usepackage{subfigure}
\usepackage{txfonts}
\usepackage{natbib}
\usepackage{url}
\usepackage{helvet}
\usepackage{setspace}
\usepackage[colorlinks=true,citecolor=blue]{hyperref}

\bibpunct{(}{)}{;}{a}{,}{,}

\def\apjl{{ApJ}}% Astrophysical Journal, Letters
\def\apjs{{ApJS}}% Astrophysical Journal, Supplement
\def\aap{{A\&A}}% Astronomy and Astrophysics
\title[SN~2013ej: Polarimetry]{Broad band polarimetric investigation of Type IIP supernova 2013ej}
\author[Brajesh Kumar et al.]{Brajesh Kumar$^{1,2,3}$\thanks{E-mail: brajesh.kumar@iiap.res.in},
{S. B. Pandey}$^{2}$,
{C. Eswaraiah}$^{4,5}$ and
{K. S. Kawabata}$^{6,7}$\\
$^{1}$Indian Institute of Astrophysics, Koramangala, Bangalore 560 034, India\\
$^{2}$Aryabhatta Research Institute of Observational Sciences, Manora Peak,
Nainital 263 002, India \\
$^{3}$ Institut d'Astrophysique et de G\'{e}ophysique, Universit\'{e} de
Li\`{e}ge, All\'{e}e du 6 Ao\^{u}t 17, B\^{a}t B5C, 4000 Li\`{e}ge, Belgium \\
$^{4}$ Institute of Astronomy, National Central University, 300 Jhongda Rd, Jhongli, 
Taoyuan Country 32054, Taiwan \\
$^{5}$ Institute of Astronomy, National Tsing Hua University (NTHU), Hsinchu 30013, Taiwan\\
$^{6}$ Hiroshima Astrophysical Science Center, Hiroshima University, Kagamiyama, 
Higashi-Hiroshima, Hiroshima 739-8526, Japan\\
$^{7}$ Department of Physical Science, Hiroshima University, Kagamiyama,
Higashi-Hiroshima 739-8526, Japan\\
}
\begin{document}
\date{Accepted ------------, Received ------------; in original form ------------}
\pagerange{\pageref{firstpage}--\pageref{lastpage}} \pubyear{}
\maketitle
\label{firstpage}
%----------------------------------------------------------------------------------------------
\begin{abstract}

We present results based on follow-up observations of the Type II-plateau supernova (SN) 2013ej at 
6 epochs spanning a total duration of $\sim$37 d. The $R_{c}$-band linear polarimetric observations
were carried out between the end of the plateau and the beginning of the nebular phases as noticed 
in the photometric light curve. The contribution due to interstellar polarization (ISP) was constrained 
by using couple of approaches, i.e. based upon the observations of foreground stars lying within 
5\arcmin\, and 10$\degr$ radius of the SN location and also investigating the extinction due to the 
Milky Way and host galaxy towards the SN direction. 
Our analysis revealed that in general the intrinsic polarization of the SN is higher than the 
polarization values for the foreground stars and exhibits an increasing trend during our observations. 
After correcting the ISP of $\sim$0.6 per cent, the maximum intrinsic polarization of SN~2013ej is 
found to be 2.14 $\pm$ 0.57 per cent. 
Such a strong polarization has rarely been seen in Type II-P SNe.
If this is the case, i.e., the `polarization bias' effect is still
negligible, the polarization could be attributed to the asymmetry
of the inner ejecta of the SN because the ISP towards the SN location 
is estimated to be, at most, 0.6 per cent.

%-------------------------------------------------------------------------------
\end{abstract}

%-------------------------------------------------------------------------------
%-------------------------------------------------------------------------------

\section{Introduction}\label{sec:introduction}

Core-collapse Type II-Plateau supernovae (II-P) are the specific events which show hydrogen 
lines in their optical spectra along with a plateau like structure in the optical light curve 
\citep[see][for a review on different Types of SNe]{1997ARA&A..35..309F}.
These events mark the end stages of the lives of massive stars  
\citep[$M>$ 8\,--10 M$_{\sun}$;][]{2003ApJ...591..288H, 2009MNRAS.399..559A}.
After explosion, the hydrogen recombination wave recedes through the outer envelope and all 
the energy deposited by the shock is slowly released \citep[e.g.][]{1993ApJ...414..712P,2009ApJ...703.2205K}.
It shows nearly constant luminosity i.e. plateau phase in the light curve. 
The plateau phase ends after approximately 100 days as the thick hydrogen envelope becomes 
optically thin and consequently a sudden drop in luminosity imprints the phase of transition.
The radioactive decay of $^{56}$Co into $^{56}$Fe powers the post-plateau phase of the light 
curve, which in turn depend upon the amount of $^{56}$Ni synthesized during the explosion 
\citep*[for different evolutionary phases of the light curve, see][]{1971Ap&SS..10...28G,1977ApJS...33..515F,
1979A&A....72..287B,2007A&A...461..233U}.

The geometry of these energetic events has been studied in detail using spectropolarimetric,
imaging polarimetric techniques \citep[e.g.][]{1988MNRAS.234..937B,1991ApJ...375..264J, 
2001ApJ...553..861L,2006Natur.440..505L,2001PASP..113..920L,2006AstL...32..739C,2010ApJ...713.1363C,
2012AIPC.1429..204L,2008ARA&A..46..433W,2014MNRAS.442....2K}, and computer simulations 
\citep[][]{1991A&A...246..481H,2006ApJ...651..366K,2011MNRAS.410.1739D}.
The first quantitative polarization study of SNe atmosphere was conducted by 
\citet[][see also \citet{1970ApJ...160.1083S}]{1982ApJ...263..902S}. 
The observed degree of polarization in these SNe may vary during various evolution phases, 
indicating the change in SN geometry and/or ejecta (e.g. density and ionization) but the 
intrinsic polarization has been found only up to $\sim$1.5 per cent \citep{2000ccsg.book.....W,
2001ApJ...553..861L,2001PASP..113..920L,2010ApJ...713.1363C}.      
It is interesting to note that although in volume-limited studies of nearby CCSNe, a large fraction 
(around 50 per cent) belongs to II-P SNe \citep[see][]{2013MNRAS.436..774E,2011MNRAS.412.1522S}, 
their polarization studies comparatively remain quite small. Multi-epoch spectropolarimetric
studies of Type II-P SN~1999em and SN~2004dj are available in the literature. \citet{2001ApJ...553..861L} 
found that polarization in SN~1999em jumped from $\sim$0.2 per cent (day 7) to $\sim$0.5 (day 160) but 
without significant change in polarization angle. In a similar study of the nearby (distance $\sim$3.13Mpc) 
SN~2004dj, \citet{2006Natur.440..505L} observed no detectable intrinsic polarization during 
the plateau phase, but there was a sudden jump up to $\sim$0.6 per cent at the end of plateau phase
as the photosphere enters the core ejecta. 
However, it is interesting to mention that in contrary to SN~1999em, SN~2004dj displayed 
rotation in polarization angle, a possible signature of the clumpy nature of Ni distribution
\citep{2006Natur.440..505L}. From the above studies, it could be inferred that despite the 
similarity in the photometric and spectral features, II-P SNe show diverse nature in their 
polarization properties.

Polarimetric follow-up observations of various II-P SNe can provide very useful 
information about these events such as the geometry of the ejecta, circum-stellar material (CSM), 
the shape of the progenitor and the explosion mechanism, etc. 
An additional advantage of the polarimetric study of II-P SNe is 
in the use of these objects as extra-galactic distance indicators 
through the expanding photosphere method \citep*[EPM;][]{1974ApJ...193...27K,1996ApJ...466..911E,
2005A&A...439..671D,2009ApJ...696.1176J,2014ApJ...782...98B}. 
The basic assumption of the EPM technique is a spherically symmetric flux distribution during 
the early stage i.e. the plateau phase. 
However, as shown by \citet{2001ApJ...553..861L}, 10 per cent asphericity may produce an EPM 
distance that overestimates the actual distance by $\sim$5 per cent for an edge-on view and 
underestimates it by $\sim$10 per cent for a face-on line of sight. 
Considering the opportunity of a bright \citep[apparent $V$ magnitude at maximum light 
$\sim$12.5 mag;][]{2014JAVSO..42..333R} nearby SN~2013ej in NGC~628, here we present the Cousins 
$R_c$-band imaging polarimetric study of this object. 

%------------------------------------------------------------------------------------------------------
\subsection{NGC~628 and its observed CCSNe}\label{ngc}

NGC~628 (also known as Messier~74; M\,74) is an interesting nearby galaxy situated in
the Pisces constellation. It is a face-on spiral galaxy (prototype SAc) with prominent 
spiral arms and dust lanes. In the sky, NGC~628 is located far away from the 
Galactic disk and therefore, is a natural target for the multi-wavelength observations 
\citep[cf.][and references therein]{2002ApJ...572L..33S, 2005ApJ...630..228K, 2006MNRAS.371.1617A,
1994ApJ...426..553C,2005ApJS..159..214K}.   
The study of star formation scenario and supernova remnants in this galaxy are also available in 
literature \citep[e.g.][]{2006ApJ...644..879E, 2000AJ....120.1306L, 2010A&A...517A..91S}.  
 
Within a distance of 10 Mpc \citep[for latest distance estimation, see][and references therein]
{2014ApJ...792...52J,2015ApJ...807...59H}, NGC~628 is one of the nearby galaxies which interestingly 
hosted three CCSNe (i.e. SN~2002ap, SN~2003gd and SN 2013ej). SN~2002ap was discovered on 2002
January 29 \citep{2002IAUC.7810....1N} and soon classified as a Type Ic event. Due to the specific 
broad spectral features, it was further recognized as a hypernova\footnote{These are highly 
energetic explosions having kinetic energy of the order of 10$^{52}$ ergs 
\citep{1998Natur.395..672I}. Few examples are SN~1997ef \citep{2000ApJ...534..660I},
SN~1998bw \citep{1998Natur.395..672I,2006ApJ...645.1331M}, SN~2003dh \citep{2003ApJ...599L..95M},
SN~2003lw \citep{2006ApJ...645.1323M}.} \citep{2002IAUC.7811....1K,2002IAUC.7811....2M,
2002IAUC.7811....3G,2002IAUC.7825....1F}. Along with detail optical spectroscopic and photometric
study \citep[c.f.][]{2002ApJ...572L..61M,2002MNRAS.332L..73G,2002ApJ...577L..97K,2003PASP..115.1220F,
2003MNRAS.340..375P,2003BASI...31..351P,2004A&A...427..453V}, this event was also monitored in 
different wavelengths e.g. radio \citep{2002ApJ...577L...5B,2006ApJ...638..930S}, X-rays 
\citep{2003A&A...397.1011S, 2004A&A...413..107S} and UV \citep{2002ApJ...572L..33S}. 
Within less than one and half year after the discovery of SN~2002ap, another supernova 
SN~2003gd was discovered by R. Evans and N.S.W. Hazelbrook in a southern spiral arm of 
NGC~628 \citet{2003IAUC.8150....2E}. The photometric and spectroscopic study of this event
was performed by \citet{2005MNRAS.359..906H} despite of its late discovery (close to the 
end of plateau phase).  
It is notable that apart from NGC~628, there are a few other galaxies such as 
NGC~6946, Arp~299, NGC~4303, NGC~5236 where 6 or more SNe have been discovered 
\citep[e.g.][]{2013A&A...550A..69A,2011MNRAS.416..567A}.

%-----------------------------------------------------------------------------------------------------
\subsubsection*{\bf SN~2013ej}

SN~2013ej is the third known SN in NGC~628 which was discovered by Lick Observatory
Supernova Search \citep[LOSS,][]{2000AIPC..522..103L} on 2013 July 25.45 \textsc{ut} 
using the 0.76m robotic Katzman Automatic Imaging Telescope (KAIT). The SN was located 
92.5 arcsec east and 135 arcsec south of the core of the host galaxy with coordinates
$\alpha = 01^{\rm h} 36^{\rm m} 48\fs16$, ~$\delta =+15\degr 45\arcmin 31\farcs0$ 
\citep{2013CBET.3606....1K}. The spectra taken on July 27.7 \textsc{ut} by 
\citet{2013CBET.3609....1V} using FLOYDS spectrograph led to the classification of 
the transient as a young Type II SN. The explosion date of the SN has been estimated with 
a precision of one day. In our study we consider the shock breakout date 
JD~2456497.45 \citep{2014MNRAS.438L.101V} as the time of explosion. 

Based on archival pre-explosion images from {\it Hubble Space Telescope} and Gemini
telescope, \citet{2014MNRAS.439L..56F} have studied the progenitor's properties. They 
provided a mass range of 8--16 M$_{\sun}$ by assuming that the progenitor
star was a red super-giant. Furthermore, this event was monitored by several groups.
\citet{2014MNRAS.438L.101V} and \citet{2014JAVSO..42..333R}, respectively presented
the initial two weeks photometry ($UBVRI/griz$, $swift-uvot$) and about 6 months photometry
($BVRI$). Similarly \citet{2015ApJ...806..160B} and \citet{2015ApJ...807...59H} have analysed 
the $UBVRI$, $swift-uvot$ and near-infrared photometry along with optical spectroscopy.  
In addition, \citet{2013ATel.5275....1L} reported earliest spectropolarimetry 
observations (taken on August 1.35 \textsc{ut}) obtained with the ESO Very Large Telescope (VLT). 
Their analysis revealed SN~2013ej to be an usual event showing strong polarization 
at a very early phase ($\sim$1.3 per cent at 430 nm to $\sim$1.0 per cent at 920 nm).

\vspace{0.2cm}
In this paper, we investigate the polarimetric properties of SN~2013ej using the 
$R_c$-band imaging polarimetry.
In Section~\ref{obs}, we present the observations and data analysis. Estimation of intrinsic 
polarization is described in Section~\ref{obsres} and finally, we discuss the results and 
summarize our conclusions in Section~\ref{conc}. 

%---------------------------------------------------------------------------------------------
\begin{figure}
\centering
\includegraphics[scale = 0.212]{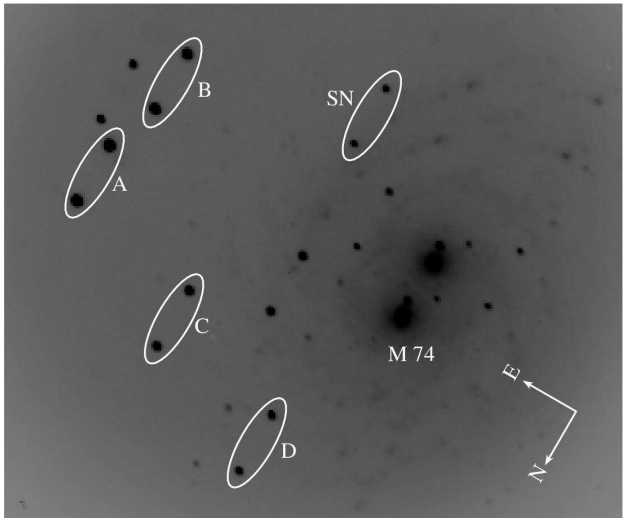}
\caption{SN~2013ej in the host galaxy NGC~628 (M\,74). The $R_c$-band image taken from 
104 cm ST on 2013 November 04 is shown. The field of view is roughly 7 arcmin 
$\times$ 6 arcmin. Each object has two images i.e. ordinary and extra-ordinary. 
Five white elliptical curves labelled with SN and letters A to D indicate the 
ordinary and extra-ordinary images of the SN~2013ej and four foreground stars 
in the field, respectively. The spiral patterns and bright knots of the galaxy 
are clearly visible. North and East directions are also indicated.}
\label{fig_1}
\end{figure}
%---------------------------------------------------------------------------------------------

\begin{table*}
\centering
\caption{Log of polarimetric observations and estimated polarization parameters of SN~2013ej. 
The $R_c$-band magnitudes listed in the last column are estimated from AIMPOL observations.
\label{tab_log}}
\begin{tabular}{ccc|cc|cc|cc}
\hline \hline
\textsc{ut} Date &JD      & Phase$^{a}$ & \multicolumn{2}{c}{Observed}  & \multicolumn{2}{c}{Intrinsic (ISP subtracted)} & Magnitude  \\
(2013)           &2456000 &(Days)       & $P_{R} \pm \sigma_{P_{R}}$    & $ \theta{_R} \pm \sigma_{\theta{_R}}$ & $P_{int} \pm \sigma_{P_{int}}$ & $\theta_{int} \pm \sigma_{\theta_{int}}$ & $R \pm \sigma_{R}$ \\
                 &        &             & (per cent)                        & ($^\circ)$                            &  (per cent)                        &  ($^\circ$)                              &     (mag)        \\
\hline                   
Oct 30 &596.09&99.54  &0.85 $\pm$ 0.32 &109.10 $\pm$ 10.68 & 0.21 $\pm$ 0.32 &112.05 $\pm$ 42.77 & 14.00 $\pm$ 0.02\\
Nov 01 &598.10&101.55 &0.65 $\pm$ 1.07 &104.58 $\pm$ 46.90 & 0.08 $\pm$ 1.07 & 66.53 $\pm$   --  & 14.41 $\pm$ 0.02\\
Nov 04 &601.06&104.51 &1.65 $\pm$ 0.29 &116.73 $\pm$  4.97 & 1.05 $\pm$ 0.29 &121.89 $\pm$  7.77 & 14.93 $\pm$ 0.03\\
Nov 10 &607.24&110.69 &1.95 $\pm$ 0.61 & 89.17 $\pm$  9.12 & 1.49 $\pm$ 0.61 & 81.56 $\pm$ 11.85 & 15.25 $\pm$ 0.03\\
Dec 04 &631.13&134.58 &2.75 $\pm$ 0.57 &115.65 $\pm$  5.85 & 2.14 $\pm$ 0.57 &117.86 $\pm$  7.58 & 15.68 $\pm$ 0.04\\
Dec 05 &632.11&135.56 &1.45 $~~~$ --  & 75.70 $~~~$ --   & 1.31 $~~~$ --  & 62.62 $~~~$ --   & 15.61 $~~~$ -- \\
\hline
\end{tabular}  \\
$^{a}$ With reference to the date of explosion JD 2456497.45. The estimates for the last epoch of observation 
indicate the limiting values. \\
\end{table*}
%-------------------------------------------------------------------------------------------------------------------

\begin{table*}
\centering
\caption{Log of polarimetric observations of field stars within 5 arcmin radius of SN~2013ej location.
The letters A, B, C and D indicates the star IDs as shown in the Fig.~\ref{fig_1} and 
IDs inside the parenthesis belongs to 2MASS.
\label{tab_log_field}}
\scriptsize
\begin{tabular}{ccc|cc|cc|cc|cc}
\hline \hline
\textsc{ut} Date &JD          & Phase$^{a}$ & \multicolumn{2}{c}{A$^{\dagger}$ (01365863+1547463)} & \multicolumn{2}{c}{B$^{\dagger}$ (01365760+1546218)}  & \multicolumn{2}{c}{C$^{\dagger}$ (01365154+1548473)}      &  \multicolumn{2}{c}{D$^{\dagger}$ (01364487+1549344)}  \\
         (2013)  &2456000     &(Days)       & $P_{R} \pm \sigma_{P_{R}}$   & $ \theta{_R} \pm \sigma_{\theta{_R}}$ & $P_{R} \pm \sigma_{P_{R}}$ & $\theta{_R} \pm \sigma_{\theta{_R}}$                           & $P_{R} \pm \sigma_{P_{R}}$                 & $ \theta{_R} \pm \sigma_{\theta{_R}}$ & $P_{R} \pm \sigma_{P_{R}}$   &$ \theta{_R} \pm \sigma_{\theta{_R}}$              \\
                 &            &             & (per cent)                &($^\circ)$            &   (per cent) &  ($^\circ$)     &   (per cent)    &($^\circ)$ &   (per cent) &  ($^\circ$)                       \\
\hline
Oct 30 &596.09&99.54  &0.69 $\pm$ 0.05& 112.18 $\pm$ 2.18 &0.87 $\pm$ 0.08& 106.55 $\pm$ 2.64 &0.44 $\pm$ 0.08&  99.21 $\pm$ 5.07 &0.24 $\pm$ 0.16& 109.31 $\pm$ 18.83 \\ 
Nov 01 &598.10&101.55 &0.82 $\pm$ 0.12& 113.39 $\pm$ 4.33 &0.75 $\pm$ 0.10& 108.08 $\pm$ 3.63 &0.65 $\pm$ 0.01& 108.09 $\pm$ 0.03 &0.39 $\pm$ 0.52&  63.02 $\pm$ 37.84 \\
Nov 04 &601.06&104.51 &0.63 $\pm$ 0.16& 112.08 $\pm$ 7.08 &0.56 $\pm$ 0.10& 107.63 $\pm$ 5.15 &0.74 $\pm$ 0.07& 111.88 $\pm$ 2.84 &0.64 $\pm$ 0.09&  99.76 $\pm$  3.95 \\
Nov 10 &607.24&110.69 &0.66 $\pm$ 0.01& 113.19 $\pm$ 0.59 &0.50 $\pm$ 0.01& 107.14 $\pm$ 0.33 &0.58 $\pm$ 0.08& 109.34 $\pm$ 3.77 &0.34 $\pm$ 0.12& 124.38 $\pm$  9.84 \\
Dec 04 &631.13&134.58 &0.67 $\pm$ 0.02& 108.86 $\pm$ 0.72 &0.54 $\pm$ 0.02& 107.57 $\pm$ 1.12 &0.56 $\pm$ 0.09& 110.73 $\pm$ 4.63 &0.14 $\pm$ 0.34& 129.81 $\pm$ 70.83 \\
Dec 05 &632.11&135.56 &0.47 $\pm$ 0.01& 108.12 $\pm$ 0.28 &0.63 $\pm$ 0.08& 102.45 $\pm$ 3.76 &0.60 $\pm$ 0.04& 111.58 $\pm$ 2.10 &0.70 $\pm$ 0.30& 104.36 $\pm$ 12.33 \\
\hline
Weighted mean of &   &   & 0.51 $\pm$ 0.01  & 108.84 $\pm$ 0.25  & 0.51 $\pm$ 0.01 & 107.14 $\pm$ 0.32 & 0.65$^{b}$ & 108.09 $\pm$ 0.04& 0.45 $\pm$ 0.06 &105.36 $\pm$ 3.94  \\
polarization parameters &&&&&&&&&&\\
\hline
\end{tabular}  \\
$^{\dagger}$ 12.16 $\pm$ 0.02, 12.58 $\pm$ 0.02, 13.51 $\pm$ 0.02 and 13.64 $\pm$ 0.02 mag are the $R_c$-band magnitudes of 
the foreground stars A, B, C and D as estimated from the AIMPOL observations. \\
$^{a}$ With reference to the date of explosion JD 2456497.45. \\
$^{b}$ Weighted mean error is less than 0.002 per cent. \\ 
\end{table*}
%-------------------------------------------------------------------------------------------------------------------

\begin{table*}
\centering
\caption{Observational detail of 9 isolated foreground stars within 10$\degr$ radius around SN~2013ej 
selected to estimate the interstellar polarization. Among them, 5 stars have known distances from
\citet{2007A&A...474..653V} catalogue.  
\label{tab:field_stars}}
\begin{tabular}{lcccll}
\hline \hline
Star      &  RA (J2000) & Dec (J2000)& $P_{R} \pm \sigma_{P_{R}}$ & $ \theta{_R} \pm \sigma_{\theta{_R}}$ &  Distance  \\
ID        &   (h:m:s)   &   (d:m:s)  &   per cent                     & ($^\circ$)                        &  (in pc )  \\ \hline
HD 11636 & 01:54:38.35 &+20:48:29.5 &0.06 $\pm$ 0.05 &79.54  $\pm$ 23.20& 17.99  $\pm$ 0.19  \\  
HD  9270 & 01:31:29.00 &+15:20:44.9 &0.08 $\pm$ 0.01 &50.59  $\pm$  1.71& 107.18 $\pm$ 8.27  \\   
HD 13248 & 02:09:46.57 &+13:10:32.9 &0.14 $\pm$ 0.06 &128.56 $\pm$ 12.32& 137.93 $\pm$ 11.23 \\  
HD  6815 & 01:08:55.85 &+09:43:49.8 &0.45 $\pm$ 0.03 &108.97 $\pm$  1.77& 167.79 $\pm$ 14.64 \\  
HD 10894 & 01:47:09.10 &+10:50:39.1 &0.19 $\pm$ 0.04 &117.03 $\pm$  6.31& 280.11 $\pm$ 33.74 \\  
HD  9560 & 01:33:55.16 &+08:39:47.5 &0.42 $\pm$ 0.02 &137.69 $\pm$  1.07& 437$^{\dagger}$    \\  
HD  8919 & 01:28:06.01 &+06:01:11.3 &0.29 $\pm$ 0.06 &117.93 $\pm$  5.84& 525$^{\dagger}$    \\ 
HD 12504 & 02:02:51.42 &+11:22:19.9 &0.18 $\pm$ 0.01 &152.77 $\pm$  0.96&  --                \\  
HD  9946 & 01:37:18.66 &+10:25:40.4 &0.11 $\pm$ 0.01 &105.28 $\pm$  3.88&  --                \\  
\hline
\end{tabular}  \\
$^{\dagger}$ These distances are adopted from \citet{2002ApJ...580L..39K}.
\end{table*}

\section{Observations and data analysis}\label{obs}

We collected broad band polarimetric data of SN~2013ej at six epochs: 2013 
October 30; November 01, 04, 10 and December 04, 05 as listed in Table~\ref{tab_log}.
These observations were conducted using the ARIES Imaging Polarimeter 
\citep*[AIMPOL,][]{2004BASI...32..159R} mounted at the Cassegrain focus 
of the 104-cm Sampurnanand telescope (ST) at Manora Peak, Nainital. 
This polarimeter consists of a half-wave plate (HWP) 
modulator and a Wollaston prism beam-splitter. The Wollaston prism 
analyzer is placed at the backend of the telescope beam path in order
to produce ordinary and extraordinary beams in slightly different
directions separated by 28 pixels along the north–south direction on
the sky plane. A focal reducer (85 mm, f/1.8) is placed between
the Wollaston prism and the CCD camera. This camera consists of 
Tektronix 1024 pixels $\times$ 1024 pixels and its cooling is performed 
by liquid nitrogen. Each pixel of the CCD corresponds to 1.73 arcsec 
and the field-of-view (FOV) is $\sim$8 arcmin in diameter 
on the sky. The full width at half-maximum (FWHM) of the stellar images vary 
from 2 to 3 pixels. The readout noise and the gain of the CCD are 
7.0 $e^{-}$ and 11.98 $e^{-}$/ADU, respectively.

Our observations were carried out in standard $R_{c}$ ($\lambda_{R_{eff}}$ 
= 0.67$\mu$m) photometric band using only the central 325 pixels $\times$ 
325 pixels of the CCD. 
Fig.~\ref{fig_1} shows approximately 7\arcmin\ $\times$ 6\arcmin\, $R_c$-band
image acquired with the AIMPOL containing SN~2013ej in the host galaxy NGC~628.
Four field stars marked with `A', `B', `C' and `D' were later used to constrain
the interstellar polarization (see Section~\ref{isp_cor}).  
Multiple frames were secured at each position angle of the HWP 
(i.e. 0$\degr$, 22.5$\degr$, 45$\degr$ and 67.5$\degr$).
The typical individual exposure time was between 10-20 minutes at
a particular position angle. 
To obtain good signal-to-noise ratio, all images at a given position angle 
were aligned and subsequently combined.

The dual-beam polarizing prism allows us to measure the polarization by simultaneously imaging 
both orthogonal polarization states onto the detector. Both the half-wave plate fast axis and 
the axis of the Wollaston prism are kept normal to the optical axis of the system. 
For such polarimeters, various Stokes parameters can be derived using standard procedures
given elsewhere \citep[e.g.][]{2006PASP..118..146P}.
In the present analysis, fluxes of ordinary and extra-ordinary beams of the SN and field 
stars were extracted by standard aperture photometry after preprocessing 
using the {\small IRAF}\footnote{{\small IRAF} is the Image Reduction and Analysis 
Facility distributed by the National Optical Astronomy Observatories, which are operated by the 
Association of Universities for Research in Astronomy, Inc., under cooperative agreement with 
the National Science Foundation.} package. 
The flux values were measured using aperture photometry at multiple apertures ranging 
from 2 to 8 pixels. These fluxes were used to derive Stokes parameters at each aperture 
\citep[c.f. equation~1,][]{2014MNRAS.442....2K}. Further, at each aperture, $P$ and $\theta$ 
values were derived by fitting the Stokes parameters with the relation $R(\alpha)=P\cos(2\theta-4\alpha)$. 
Where, $P$ and $\theta$ are degree of polarization and polarization angle, $\alpha$ is the position angle 
(0$^\circ$, 22.5$^\circ$, 45$^\circ$ and 67.5$^\circ$) of HWP. Final $P$  and $\theta$ values were 
chosen at the aperture that best fit with minimum chi-square. Generally, best fit has always been witnessed 
between 4 to 6 pixels corresponding to 2 to 3 FWHM. Table~\ref{tab_log} provides the estimated polarization 
measurements. 
The observed degree of polarization and polarization angle are denoted as $P_R$ (per cent) and
$\theta_{R}$ ($^\circ$), respectively.

%------------------------------------------------------------------------------------------
\begin{table}
\centering
\caption{ Results of observed polarized standard stars.
\label{stand_log}}
\begin{tabular}{lcccc}
\hline \hline
Date of observation  &  $P_{R} \pm \sigma_{P_{R}}$  & $\theta{_R} \pm \sigma_{\theta{_R}}$  \\
(\textsc{ut} 2013)   &         (per cent)           &            ($^\circ)$                 \\
\hline
HD~19820 (Std. values) & 4.53 $\pm$ 0.03           & 114.46 $\pm$ 0.16                     \\
Oct 30                 &  4.6  $\pm$ 0.1           & 114    $\pm$ 1                        \\ \\
HD~59389 (Std. values) & 6.43 $\pm$ 0.02            & 98.14  $\pm$ 0.10                     \\
Nov 01                 & 6.3  $\pm$ 0.1             & 98     $\pm$ 1                        \\ \\
HD~25443 (Std. values) & 4.73 $\pm$ 0.32            & 133.65 $\pm$ 0.28                     \\
Nov 04                 & 4.8  $\pm$ 0.1             & 133    $\pm$ 2                        \\
Nov 10                 & 4.6  $\pm$ 0.2             & 133    $\pm$ 2                        \\
Dec 04                 & 4.9  $\pm$ 0.2             & 134    $\pm$ 1                        \\
Dec 05                 & 4.9  $\pm$ 0.1             & 134    $\pm$ 2                        \\ \\
HD~43384 (Std. values) & 2.86 $\pm$ 0.03& 170.7  $\pm$ 0.7                      \\
Nov 10                 & 2.8  $\pm$ 0.1             & 170    $\pm$ 1                        \\ \\

HD~23663 (Std. values) & 5.38 $\pm$ 0.03            & 93.04  $\pm$ 0.15                     \\
Dec 04                 & 5.4  $\pm$ 0.1             & 93     $\pm$ 1                        \\
Dec 05                 & 5.3  $\pm$ 0.1             & 94     $\pm$ 2                        \\
\hline
\end{tabular}  \\
All standard values listed here are from \citet*{1992AJ....104.1563S} except HD~43384 which is
taken from \citet{1982ApJ...262..732H}. \\
\end{table}
%-----------------------------------------------------------------------------------------

To correct our measurements for the zero-point polarization angle, we performed 
observations of several polarized standards stars taken from \citet*{1992AJ....104.1563S}
and \citet{1982ApJ...262..732H}.
The results are given in Table~\ref{stand_log} which indicate a good agreement (within the 
observational errors) with standard values.
The difference between the observed and standard values were applied to the SN for the 
respective dates.
It is worth mentioning that the instrumental polarization of focal reducers is known to 
grow from the optical axis to the edges of the field-of-view \citep[][]{2006PASP..118..146P}. 
Nevertheless, the SN was observed slightly off-axis which yielded simultaneous imaging
of few field stars along with the SN (see also Section~\ref{fg}). 
The instrumental polarization of AIMPOL on the 104 cm ST has been characterized and monitored 
since 2004 for different projects and generally found to be $\sim$0.1 per cent in $R_c$-band
\citep[e.g.,][and references therein]{2004BASI...32..159R, 2009MNRAS.396.1004P, 2011MNRAS.411.1418E,
2013A&A...556A..65E,2014MNRAS.442....2K,2015A&A...573A..34S}. 
To verify this result, we further observed unpolarized standards stars G~191\,B2B and HD~21447 
from \citet*{1992AJ....104.1563S} on different nights and their degree of polarization was 
estimated to be 0.13 $\pm$ 0.03 and 0.14 $\pm$ 0.05 per cent, respectively. However, in our 
analysis we have applied an average correction of 0.1 per cent to the observed polarization 
values of the SN. 

%------------------------------------------------------------------------------------------
 
It is worthwhile to note that present set of images were obtained in absence of a 
window mask on the focal plane. In such a set-up, the polarization study of well 
isolated bright stars normally do not impose any problem. 
But it may be an issue for the objects which are lying over diffuse backgrounds
(e.g. SNe) because the background polarization is not removed properly.
A variable amount of polarization may also be introduced due to different phases 
of moon and/or seeing effects. 
In our analysis we made an attempt to quantify the possible effect of contamination 
by the nebulous background on the final polarization measurements of SN~2013ej. 
A detailed description of the method is given in Appendix~\ref{apn_a}. 
Our exercise suggests that although there is a non-negligible nebulous 
background present around SN~2013ej, it does not significantly influence the final 
polarization measurements.

%-----------------------------------------------------------------------------------------------------

\section{Observational results}\label{obsres}

\subsection{Constraining interstellar polarization}\label{isp_cor}

The interstellar polarization (ISP) is produced by differential extinction 
resulting from aspherical and spinning dust grains along the line of sight 
that are aligned in space by the Galactic magnetic field
\citep[see][]{2003dge..conf.....W}.
The ISP (either due to the Milky Way and/or host galaxy)
may largely influence the observed SN polarization properties. For example,
\citet{2001PASP..113..920L} found an extraordinarily high degree of polarization 
(5.8 per cent) for SN~1999gi in the galaxy NGC~3184 but later, in a separate study,
\citet{2002AJ....124.2506L} estimated that the previously derived value was 
heavily affected by ISP, and the intrinsic polarization of the SN was
only in the range of 0.3 to 2.0 per cent. 
Therefore, to understand the intrinsic behavior of the SN polarization parameters, 
it requires careful accounting for the intervening interstellar medium. 
Due to unavailability of any standard method for the estimation of ISP, every
possible measure should be considered to avoid any observational biases.
Fortunately, ISP has been well studied in the Galaxy \citep*[e.g.][]{1975ApJ...196..261S}. 
Furthermore, while SN polarization properties may show temporal variation, the ISP 
should be constant. Therefore, by constraining the ISP we can subtract it from the 
observed SN polarization parameters to obtain intrinsic polarization measurements of SN. 

As mentioned in Section~\ref{ngc}, along with SN~2013ej, NGC~628 had also hosted 
two additional CCSNe (c.f. SN~2002ap and SN~2003gd). 
Contrary to SN~2003gd, early discovery and brightness of SN~2002ap
\citep[peak apparent $V$ mag $\sim$12.4,][]{2003MNRAS.340..375P} provided better 
opportunity to study the spectropolarimetric properties in great details by several
groups. These authors have also discussed about the ISP contributions. 
\citet{2002PASP..114.1333L} examined ISP for SN~2002ap using the $q-u$ plane 
method, Serkowski law ISP curve fitting method and by measuring the polarization
of five distant stars within 1$\degr$ of the line of sight of SN. There was 
general agreement in all estimated values and they adopt $p_{isp}$ = 0.51 per cent at
$\theta_{isp}$ = 125$\degr$. Similarly \citet{2002ApJ...580L..39K} and 
\citet{ 2003ApJ...592..457W} have estimated ISP as $p_{isp}$ = 0.64 per cent, 
$\theta_{isp}$ = 120$\degr$; $p_{isp}$ = 0.61 per cent, $\theta_{isp}$ = 122$\degr$, 
respectively. 

The ISP characterization in the above cited studies provide us an insight about the 
dust properties in the direction of NGC~628 and enable us to evaluate intrinsic 
polarization properties of SN~2013ej. In the following sections, we present various 
approaches which we applied to obtain the intrinsic polarization measurements of SN~2013ej.
%-------------------------------------------------------------------------------------

\begin{figure*}
\begin{centering}
\includegraphics[scale=0.4357]{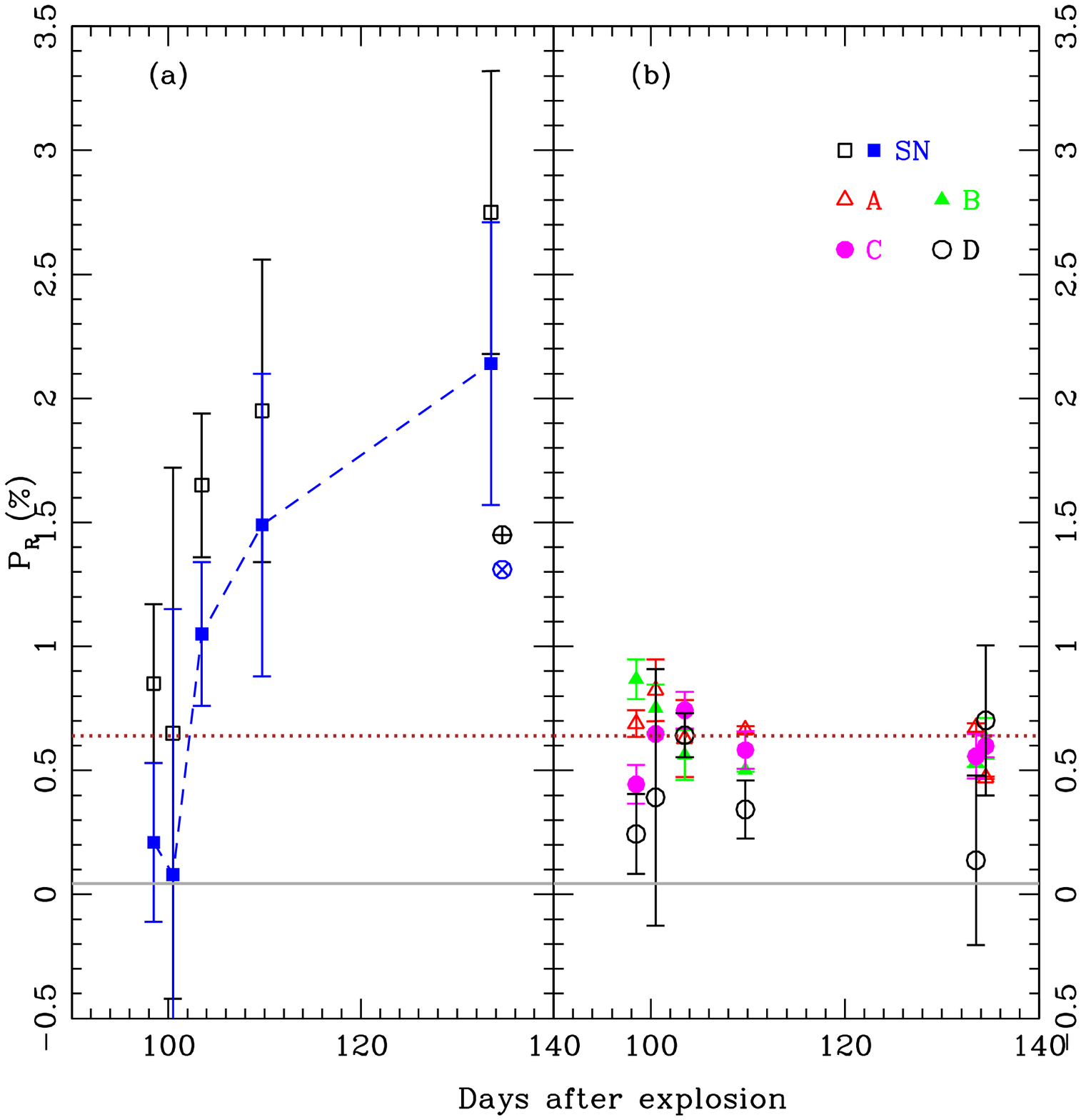}
\includegraphics[scale=0.4357]{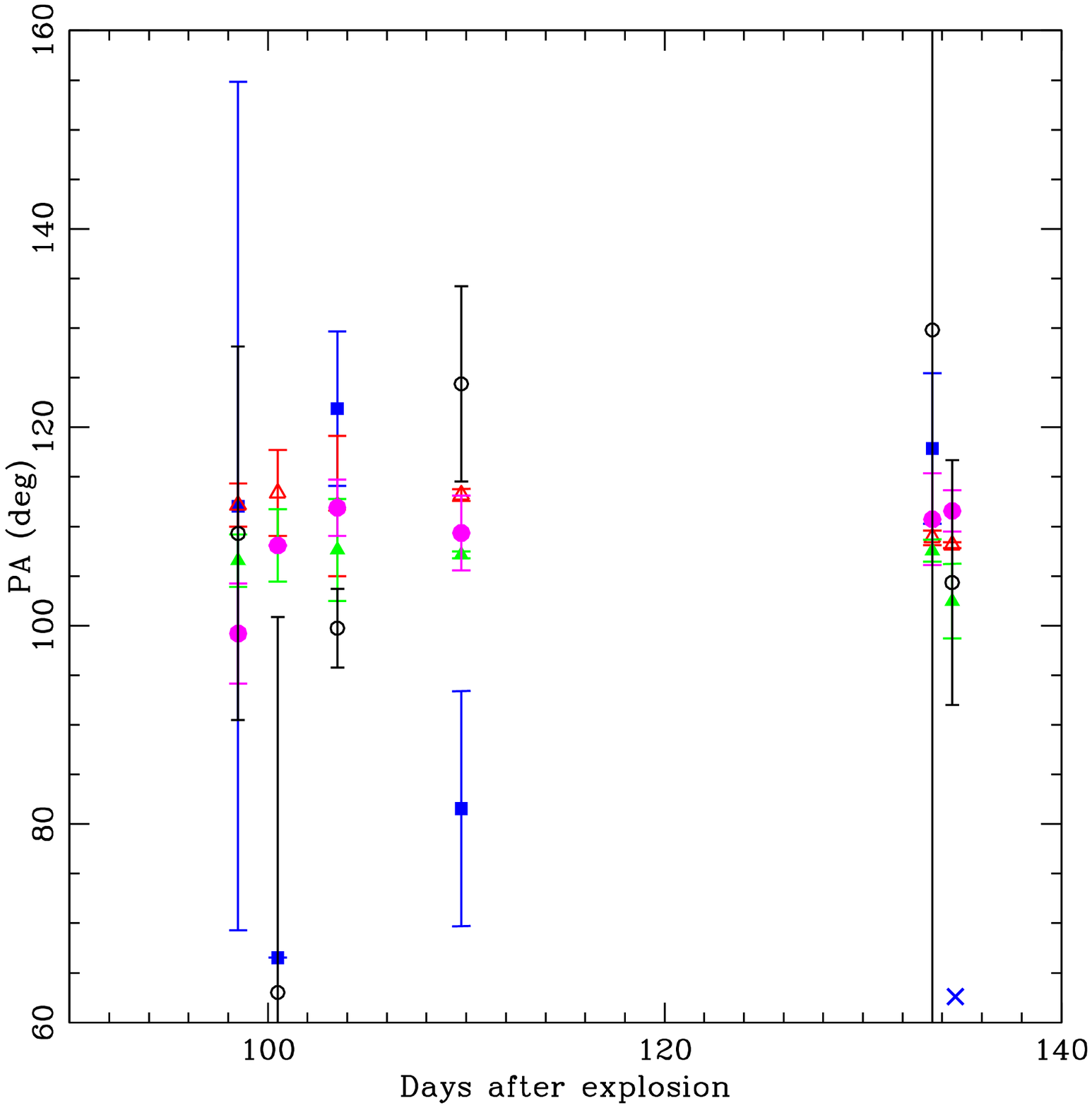}
\caption{Temporal evolution of polarization parameters of SN~2013ej. {\bf Left panel}: Evolution of $P_{R}$
values of SN and foreground stars (lying within 5 arcmin radius of SN). The observed (open square)
and intrinsic (filled square and connected with dashed line) $P_{R}$ values of SN (in panel-a) and
observed $P_{R}$ values of foreground stars (in panel-b) are shown with different symbols. The dotted
line (brown colour) and solid line (grey colour) respectively, indicate the weighted mean of $P_{R}$
values of four foreground stars as shown in Fig.~\ref{fig_1} and weighted mean of eight foreground
stars observed with AIMPOL in $R_c$-band (see Section~\ref{tendeg} for details). The encircled plus
(black colour) and cross (blue colour) symbols (in panel-a) represent the observed and intrinsic
limiting polarization values, respectively. {\bf Right panel}: Evolution of intrinsic PA of the SN 
and foreground stars. 
The cross symbol (blue colour) represents the intrinsic limiting PA of the SN and other symbols 
in both panels have similar meaning.}
\label{p_pa_evol}
\end{centering}
\end{figure*}
%-------------------------------------------------------------------------------------

\subsubsection{Fore-ground stars lying within 5$\arcmin$ radius of SN position}\label{fg}

There are several isolated and medium brightness field stars situated within 5 arcmin
radius of the SN location. Since AIMPOL provides the 8 arcmin diameter wider field-of-view, 
we are fortunate enough to observe several field stars along with SN. Therefore, during the 
period of our observations, we pointed the telescope in such a way that all those stars were 
also observed along with SN. 
To characterize the polarization properties of these nearby stars, initially we selected
six stars with good S/N ratio. 
Out of them, the trend of two stars were consistent with zero polarization within 
3-sigma but the polarization of SN after plateau phase was significantly large.
Therefore, we excluded two stars and only the remaining four field stars were considered
for the further analysis. 
These are labelled with letters `A' to `D' as shown in Fig.~\ref{fig_1} and their estimated 
polarization parameters are listed in Table~\ref{tab_log_field} (Two Micron All-Sky 
Survey \citep[2MASS,][]{2006AJ....131.1163S} IDs are also mentioned in the parenthesis).  
%%%-----------------------------------------------------------------------------
To cross-examine the effect of growing noise on polarimetric results of SN 
(especially towards the late epochs), it could have been interesting to derive the 
magnitudes of the fainter objects in the SN field. However, some of them are situated 
in the nebulous regions so we avoided those stars. Nonetheless, $R_c$-band magnitudes 
of `A'-`D' stars are mentioned in the footnote of Table~\ref{tab_log_field}.    

%-------------------------------------------------------------------------------------

\subsubsection{Fore-ground stars lying within 10$\degr$ radius of SN position}\label{tendeg}

We examined the polarization measurements of several foreground Milky Way stars available
in the ISP catalogue by \citet{2000AJ....119..923H}. Within 10$\degr$ radius around the 
SN~2013ej, we selected only 9 stars which are isolated and do not show either 
emission features or variability flag in the SIMBAD data base. Out of them, 5 stars 
have distance information from {\it Hipparcos} parallax \citep{2007A&A...474..653V}. 
Distances of HD~8919 and HD~9560 have been adopted from \citep{2002ApJ...580L..39K}. 
Considering the fact that Heiles catalogue represents the polarization in a band different 
than ours, we have performed $R_c$-band polarimetric observations of these stars 
and their estimated $P$ and $PA$ values are listed in Table~\ref{tab:field_stars}. 
Among this list, HD~11636 is only at a distance of $\sim$18 pc thus likely to skew 
the results because it is nearby and does not probe all material in the Galactic plane.
We have excluded this star from our further analysis.
The weighted mean values of remaining 8 stars are as follows:

\vspace{0.2cm}
$<$$P_{heiles}$$>$ = 0.025 $\pm$ 0.003, $<$$PA_{heiles}$$>$ = 125.08 $\pm$ 3.22 
(for Heiles catalogue) and
$<$$P_{aimpol}$$>$ = 0.044 $\pm$ 0.004, $<$$PA_{aimpol}$$>$ = 55.82 $\pm$ 2.26
(for AIMPOL $R_c$-band), respectively.
The polarization level (weighted mean) of these eight stars ($<$$P_{aimpol}$$>$) as observed 
with AIMPOL (in $R_c$-band) is indicated by a grey continuous line in Fig.~\ref{p_pa_evol}. 

%------------------------------------------------------------------------------------

\subsubsection{Extinction due to the Milky Way and the host galaxy}\label{ext}

Based on the observations of various nearby Galactic stars, 
\citet*[][see also \citet{2003dge..conf.....W}]{1975ApJ...196..261S}
have provided an empirical relation between reddening and maximum ISP which is given as
$P_{\rm ISP_{max}}$ = 9\,$\times$\,$E(B-V)$, where $E(B-V)$ indicates the reddening value.
Although this relation is not universal and has shown some violations \citep[for example,
][and references therein]{2001PASP..113..920L,2002AJ....124.2490L,2011A&A...527L...6P} but
it is still valid for general investigation purpose.

The $E(B-V)$ resulting from Milky Way interstellar matter in the line of sight of SN~2013ej
was found to be 0.062 mag \citep{2011ApJ...737..103S} which is corresponding to an extinction 
of 0.19 mag in $V$-band for total-to-selective extinction parameter $R_V$ = 3.1 \citep{1989ApJ...345..245C}. 
\citet{2015ApJ...806..160B} have used high resolution echelle spectra (obtained on 79.5 d) 
and analysed the equivalent width of Na I D doublets to estimate the reddening in the
line of sight of SN~2013ej. They clearly resolved these lines of the Milky Way, but 
Na I D impression for the NGC~628 was missing (see their fig.~2). This implies that 
the host galaxy contribution is negligible and only Galactic reddening is dominantly 
playing a role for the SN line of sight extinction. \citet{2014MNRAS.438L.101V} have also 
concluded similar results (see their fig.~3). Now we consider total reddening in the
direction of SN $E(B-V)$ = 0.062 mag and thus, the maximum ISP relation gives the maximum 
degree of ISP towards the SN~2013ej as 0.558 per cent.

%%%-----------------------------------------------------------------------------------

\subsection{Estimation of SN intrinsic polarization}\label{isp_cal}

From the previous discussions, it is clear that MW dust is significantly contributing 
to the observed SN polarization measurements, and hence it must be subtracted.
The foreground stars lying in 10$\degr$\, radius around the SN gives us a general
indication of the polarization in this region of sky and consequently along the SN
line of sight. But, it is noteworthy to mention that polarization estimates
of such sample of stars may not provide exact magnitude and direction of the ISP
specially when these stars are beyond $\sim$1\degr\, from the SN location 
\citep[c.f.][]{1995ApJ...440..565T,2001PASP..113..920L}. 
Therefore, for the IPS contribution we focus only on those four stars which are 
within 5\arcmin\, radius of the SN. Ideally, we should know the distance (or reddening) 
and the spectral features (which may infer that they are intrinsically unpolarized or not) 
of these field stars. However, practically it is not an easy task. A careful investigation 
of Fig.~\ref{p_pa_evol} (see also Table~\ref{tab_log_field}) indicates that in the course 
of follow-up observations, polarization parameters of these stars do not show significant 
variation. This homogeneity of the polarization values among `A'-`D' stars may suggest that they 
represent uniform MW ISP in the direction of SN~2013ej.
Subsequently, we derived the weighted mean of polarization parameters of these four stars.
For each star `A', `B', `C' and `D', first we derived their individual weighted mean of 
polarization and polarization angles for each epoch as listed in Table~\ref{tab_log_field}.
Afterwards, the weighted mean of these values were estimated which are found to be
0.637 $\pm$ 0.001 per cent and 108.11 $\pm$ 0.04 degree. Let us denote it
as $<$$P_{4s}$$>$ and $<$$PA_{4s}$$>$, respectively, i.e. $<$$P_{4s}$$>$ = 0.637 $\pm$
0.001 per cent and $<$$PA_{4s}$$>$ = 108.11 $\pm$ 0.04 degree, where suffix `4s' stands for 
4 stars lying in the same field-of-view of SN (within 5\arcmin\, radius).

As mentioned previously that for SN~2002ap, \citet{2002ApJ...580L..39K},
\citet{2002PASP..114.1333L} and \citet{2003ApJ...592..457W}
have determined $P_{isp}$ as 0.64, 0.51 and 0.61 per cent, respectively. Although
detailed polarimetric study of SN~2003gd (which was also discovered in the same galaxy,
see Section~\ref{ngc}) is not available in the literature, \citet{2005ASPC..342..330L} have 
mentioned about 1 per cent ISP contribution towards the line of sight of this object. 
The $<$$P_{4s}$$>$ value (0.637 per cent) is much larger than $<$$P_{aimpol}$$>$ value 
(0.044 per cent), but it is comparable with the values derived in the previously mentioned 
studies as well as estimated from the maximum ISP and reddening relation given by 
\citet*{1975ApJ...196..261S}. Moreover, it seems that there is no noticeable extinction
at the position of SN due to host galaxy (see Section~\ref{ext}).
Above facts strongly suggest that total ISP contribution in the SN direction is not more 
than 0.637 per cent and the polarization parameters of 4 field stars around SN represent 
real ISP in the direction of SN~2013ej.  

By following the methods as described in \citet[][see their Section~3.1]{2014MNRAS.442....2K}, 
we have estimated and subtracted the ISP components from the observed SN polarization measurements. 
Firstly, observed $P_R$ and $\theta_{R}$ of SN as well as four field stars were transformed into 
Stokes parameters, then weighted mean Stokes parameters of field stars were determined as 
$<$$Q_{4s}$$>$ = -- 0.515 $\pm$ 0.001 per cent and $<$$U_{4s}$$>$ = -- 0.378 $\pm$ 0.001 per cent. 
To obtain the intrinsic Stokes parameters of SN, the $<$$Q_{4s}$$>$ and $<$$U_{4s}$$>$ values were 
vectorially subtracted from the observed Stokes parameters of SN for each epoch. These values were 
further converted back to $P_{int}$ and $\theta_{int}$ \citep[see relations 5, 6, 7 and 8 in]
[for more details] {2014MNRAS.442....2K}. The intrinsic SN~2013ej polarization values 
(denoted as $P_{int}$ and $\theta_{int}$) for each epoch are listed in column 6 and 7 in 
Table~\ref{tab_log}.

In the left panel of Fig.~\ref{p_pa_evol}, we have shown the temporal evolution of observed and 
intrinsic polarization values of SN~2013ej with open and filled square symbols, respectively. 
The polarization measurements of 4 foreground stars (within 5$\arcmin$ radius of SN) are also 
over plotted with different symbols. Two horizontal lines dotted (brown colour) and continuous 
(grey colour), respectively illustrate the level of polarization estimated for foreground stars 
within 5$\arcmin$ and 10$\degr$ radius around the SN location. In the right panel of the figure,
intrinsic polarization angles of SN and 4 field stars are over plotted.
The degree of polarization is known to be suffered from `polarization bias'
effects for errornous data having smaller $P/\sigma_{P}$ 
\citep[e.g.][]{1958AcA.....8..135S,1974ApJ...194..249W,2006PASP..118..146P}.
Adopting a traditional, statistic correction as
$P_{\rm real}=\sqrt{P^{2}-(\sigma_{P})^{2}}$, the $P$ decreases by a
factor of $0.25$ for a data point having $P/\sigma_{P}=2$.
Thus, we caution the reader that if we make corrections for 
de-bias, then our initial three estimated values of the SN will tend towards the null polarization.

%--------------------------------------------------------------------------------------------
\begin{figure}
\begin{centering}
\includegraphics[scale=0.42]{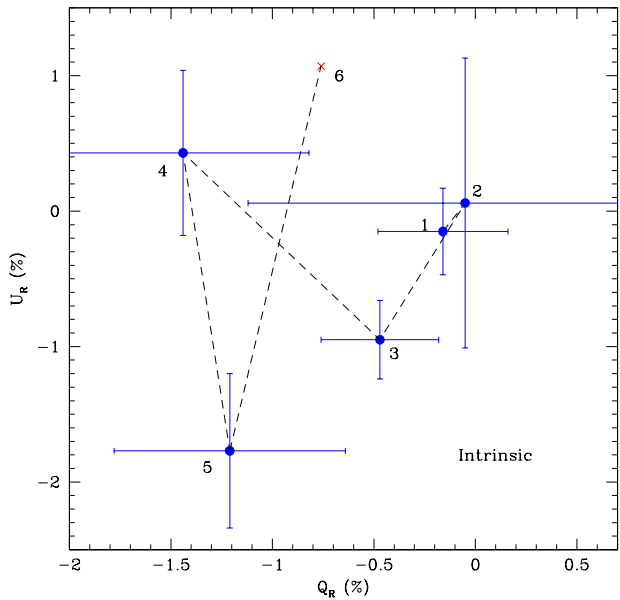}
\caption{$Q-U$ diagram for SN~2013ej. Labelled numbers (1--6) connected with dashed line
represent the temporal order. Blue coloured dots are the ISP corrected polarization data and
cross symbol (red colour) indicates the limiting value at the final epoch of observation.}
\label{qu_evol}
\end{centering}
\end{figure}

%--------------------------------------------------------------------------------------------

\begin{figure*}
\begin{centering}
\includegraphics[scale=1.42]{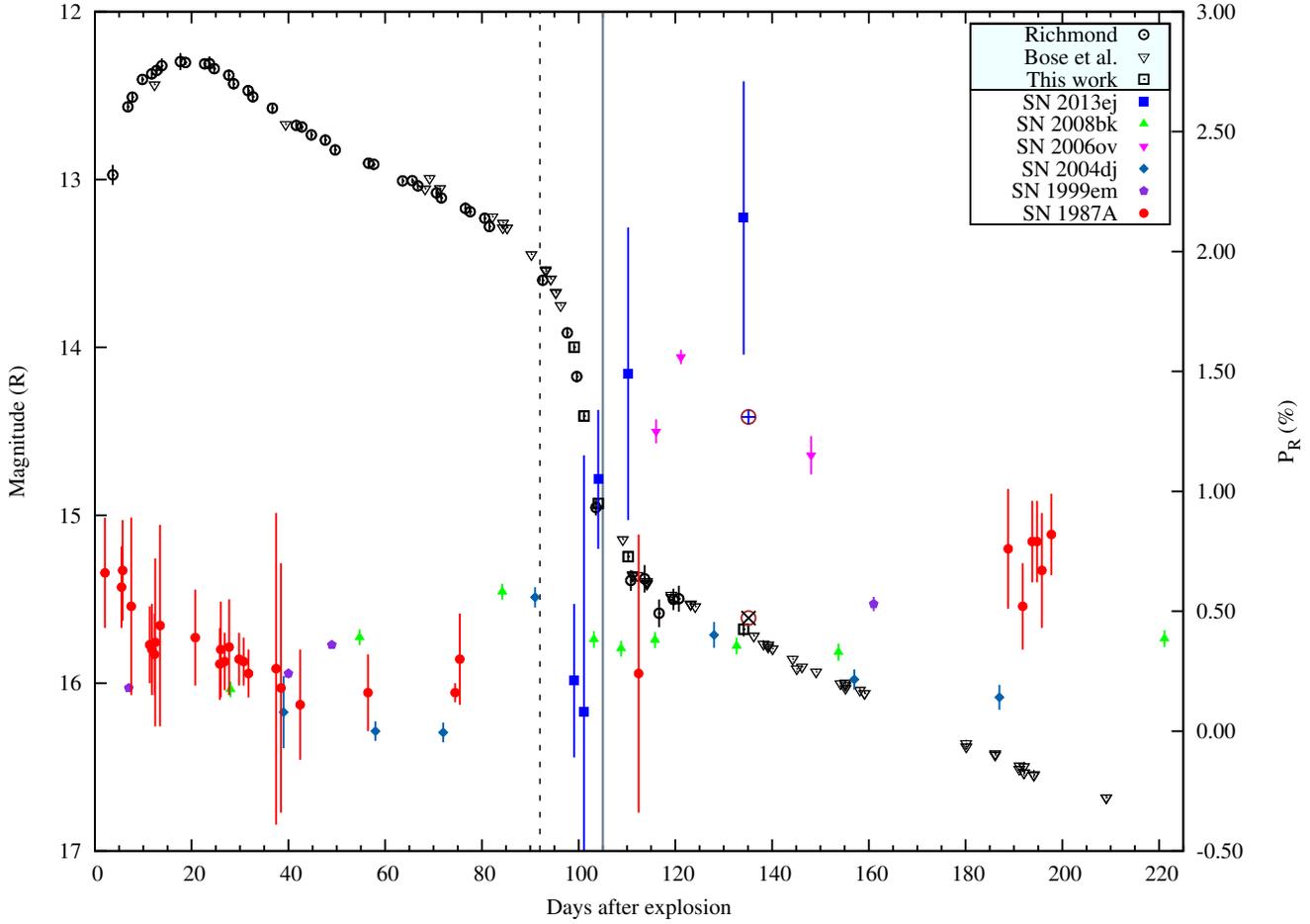}
\caption{Photometric and polarimetric evolution of SN 2013ej. The calibrated $R_c$-band photometric
light curve from \citet{2014JAVSO..42..333R}, \citet{2015ApJ...806..160B} and estimated from the
AIMPOL observations in present study, are shown with different symbols (inside light-cyan
shaded background in upper side of the legend).
The intrinsic polarization values of different Type II-P SNe (indicated in the legend) along with
SN~2013ej are over-plotted for comparison. Two vertical lines, dotted (black colour)
and bold (grey colour), respectively indicate the end of plateau and beginning of nebular
phases (see Section ~\ref{conc} for details).
The encircled (brown colour) plus and cross symbols represents the limiting polarization
and limiting photometric magnitude, respectively.}
\label{phot_pol}
\end{centering}
\end{figure*}

%-------------------------------------------------------------------------------------------------------
\begin{table*}
\centering
\caption{This table summarizes the ejected nickel mass, plateau decay rate and explosion energy
of 6 Type II-P SNe including the present study. 
References: 
(1) \citet{2015ApJ...806..160B}; 
(2) \citet{2015ApJ...807...59H};
(3) \citet{2013msao.confE.176P}; 
(4) \citet{2014ApJ...786...67A};
(5) \citet{2009MNRAS.395.1409S}; 
(6) \citet{2014MNRAS.439.2873S};
(7) \citet{2006AJ....131.2245Z};
(8) \citet{2009ApJ...695..619V};
(9) \citet{2003MNRAS.338..939E};
(10) \citet{1990AJ.....99.1133P}; 
(11) \citet{1996snih.book.....A} and
(12) \citet{2015A&A...575A.100U}.
\label{nickel}}
\begin{tabular}{ccccc}
\hline \hline
SN       &Nickel mass      &Plateau decline rate$^{a}$ &Explosion energy &References \\
         &(M$_\odot$)      &                           & (10$^{51}$ erg) & \\ \hline
SN~2013ej&0.020 $\pm$ 0.002&1.74 $\pm$ 0.08 (mag/100 d)& 0.7& 1, 2 \\             
         &0.020 $\pm$ 0.010&                           &    &      \\
SN~2008bk&0.009 $\pm$ $...$&0.11 $\pm$ 0.02 (mag/100 d)& 0.3& 3, 4 \\
         &0.007 $\pm$ 0.001&                           &    &      \\  
SN~2006ov&0.003 $\pm$ 0.002&0.53 $\pm$ 0.05 (mag/20 d) &$...$ & 5, 6 \\
         &0.002 $\pm$ 0.002&                           &    &      \\
SN~2004dj&0.023 $\pm$ 0.005&0.50 $\pm$ 0.04 (mag/20 d) & 0.8& 7, 8 \\
         &0.020 $\pm$ 0.010&                           &    &      \\
SN~1999em&0.027 $\pm$ 0.002&0.31 $\pm$ 0.02 (mag/100 d)& 1.3& 1, 2, 9 \\
         &0.050 $\pm$ 0.010&                           &    &      \\
SN~1987A &0.075 $\pm$ 0.005&      $...$                & 1.5& 10, 11, 12\\
         &0.077 $\pm$ $...$&                           &    &      \\
\hline
\end{tabular}  \\
$^{a}$ The decline rate in $V$-band.  
\end{table*}

%%%---------------------------------------------------------------------------------------------

\section{Discussion and conclusion}\label{conc}

The polarimetric follow-up observations (including spectropolarimetry and broad band)
of the Type II-P SNe are very limited, moreover the data sampling is sparse. In the literature, 
other than SN~2013ej, there are only 5 such events which have at least three epochs of 
polarization measurements and are extended up to the nebular phase. 
These include SN~2008bk \citep{2012AIPC.1429..204L},
SN~2006ov \citep{2010ApJ...713.1363C}, 2004dj \citep{2006Natur.440..505L}, 1999em
\citep{2001ApJ...553..861L} and SN~1987A \citep{1988MNRAS.234..937B}.
We have compared their polarimetric properties in Fig.~\ref{phot_pol} along with SN~2013ej.
It is worthy to mention here that except SN~2013ej, SN~1999em and SN~1987A, the explosion
dates of other events are not known so precisely.

In Fig.~\ref{phot_pol} we have shown the $R_c$-band photometric light curve of SN~2013ej
from \citet{2014JAVSO..42..333R}, \citet{2015ApJ...806..160B} and also the magnitude estimated
from AIMPOL polarimetric observations. 
To calibrate the observed AIMPOL SN magnitudes, we selected the secondary standard 
star from \citet[][see his fig.~1 where particular star is denoted by ID \#A and it is
common in both studies]{2014JAVSO..42..333R}. Then for each night observation, the offset 
values in both magnitudes were applied to the observed AIMPOL SN magnitudes.

The estimated SN~2013ej magnitudes at 6 epochs are listed in Table~\ref{tab_log} and plotted 
with open square symbols (black colour) in the Fig.~\ref{phot_pol}. As it can be seen, it is 
very nicely following the light curve evolution pattern of the other two studies 
\citep[i.e.][]{2014JAVSO..42..333R,2015ApJ...806..160B}. 
Two vertical lines in Fig.~\ref{phot_pol} mark two important phases, the end of plateau 
phase ($\sim$ 92d, with red dashed line) and beginning of nebular phase ($\sim$105 with 
continuous grey line). The intrinsic polarization values of SN ($P_{int}$) are over-plotted 
on it with filled square symbols. It is obvious from Fig.~\ref{phot_pol} that our 
polarimetric follow-up observations are very unique in the sense that these data samples 
belongs to the crucial evolutionary phase, i.e. end of plateau phase when the hydrogen envelope 
is almost recombined and inner core is revealed. Furthermore, last three data points 
were obtained just around the beginning of nebular phase.

Since the variation pattern of the $Q-U$ parameters does not depend to the ISP corrections
therefore, it may provide a better probe to examine the simultaneous behaviour of the 
SN polarization \citep[see, e.g.][and references therein]{2008ARA&A..46..433W}. 
In Fig.~\ref{qu_evol} we have plotted the ISP subtracted polarization data of SN~2013ej
on the $Q-U$ plane where each number is indicative of data point per epoch. 
It is apparent from this figure that there are large error bars in our estimates which
do not allow us to establish firm conclusion on the observed features of this event based 
on the imaging polarimetric measurements.

The morphological information about expanding SN ejecta can be directly probed by
polarimetric studies \citep{1982ApJ...263..902S,1991A&A...246..481H,2008ARA&A..46..433W}.
Spectropolarimetry provides useful information about the
overall shape of the emitting region and dynamical evolution of various chemical
elements of the explosion. However, broad-band imaging polarimetry allow us longer
follow-up coverage of bright events which may pass through important phases
of their temporal evolution. A net polarization in hot young CCSNe arises due to incomplete
cancelation of directional components of electrical vectors which finally
reveal about possible asphericity or clumpiness in the ejecta. In a spherical source, 
the condition is entirely different when these vectors exactly cancelled and yield zero 
net polarization \citep[see][]{2005ASPC..342..330L, 2008ARA&A..46..433W, 2003ApJ...593..788K}.
There are couple of other processes which may cause polarization in CCSNe such as 
asymmetric distribution of radioactive material within the SN envelope 
\citep[e.g.][]{1995ApJ...440..821H, 2006AstL...32..739C}, aspherical ionization produced 
by hard X$-$rays from the interaction between the SN shock front and a non-spherical 
progenitor wind \citep{1996ssr..conf..241W} and/or scattering by dust 
\citep[e.g.][]{1996ApJ...462L..27W}.

\citet{2010ApJ...713.1363C} have studied the nature of asphericity in three II-P SNe 
(SN~2006my, SN~2006ov and SN~2007aa) and hinted that along with several factors like
progenitor mass, explosion energy and $^{56}$Ni mass ejection may play a possible role 
towards the expected net continuum polarization. Among those compared objects, SN~2006ov 
ejected smallest amount of $^{56}$Ni 
and exhibited larger degree of polarization. We revisited their findings and for this purpose 
collected sample of explosion energy and ejected $^{56}$Ni mass of Type II-P SNe from the 
literature. These parameters are listed in Table~\ref{nickel}. 
Here, one should keep in mind that $^{56}$Ni mass estimation procedures were different in 
respective studies (i.e. hydrodynamical as well as analytical). 
From this sample it is apparent that the amount of $^{56}$Ni mass in SN~2013ej is comparable
with SN~2004dj and SN~1999em but considerably higher than SN~2006ov (more than 6 times) and
SN~2008bk (more than 2 times). 
We also estimated the plateau decay rate (PDR) of the existing sample in $V$-band light curves. 
PDR is expressed as the decay of magnitude between 100 days of the plateau phase. However,
it is worthy to mention that because of early discovery, SN~2013ej, SN~2008bk and SN~1999em 
have a good sampling of the data. Unfortunately, SN~2006ov and SN~2004dj were discovered 
at late epochs and therefore, we estimate PDR only for 20 days for these events due to 
lack of the data. Furthermore, since SN~1987A is a peculiar event, it was not considered 
to estimate PDR. Though PDR of SN~2006ov and SN~2004dj may not be good representative 
for comparison with other objects, but it may provide us a general overview about the light 
curve decay.
The estimated PDR values are also listed in Table~\ref{nickel}. Interestingly, with a
value of 1.74 mag/100 d, SN~2013ej has the highest rate of plateau decay.
Here we caution that the existing sample is too small to draw any conclusion about
the possible correlation between the ejected $^{56}$Ni mass and the plateau decay 
rate and the observed polarization evolution \citep[see also][]{2010ApJ...713.1363C}.
It should also be noted that in a recent study of multi-band model of nearby well-observed 
Type II-P SNe \citet{2015ApJ...806..225P} examine the significance of the correlations between 
the various parameters of explosion such as plateau luminosity, nickel mass, explosion energy 
and ejecta mass. They advocate that Type II-P SN explosions are governed by several physical 
parameters which in turn reflect the diversity of the core and surface properties of their 
progenitors.

As can be seen in Fig.~\ref{phot_pol}, although there are large error bars in 
polarization estimates of SN~2013ej but in general, the intrinsic polarization values 
show increasing trend during our observations and ranges between 0.08 to 2.14 per cent.
The relatively higher degree of polarization estimated from our analysis is also 
supported by the spectropolarimetric observation in the early epoch where 
\citet{2013ATel.5275....1L} have found strong polarization of about
1 per cent for this event.
The maximum intrinsic polarization reached up to a level of 2.14 $\pm$ 0.57 per cent, 
which is significantly higher than similar Type II-P events.
The $PA$ of SN~2013ej exhibits superimposed variability during the observed period, 
though affected with large error bars (see Fig.~\ref{p_pa_evol}). Such variations in 
the $PA$ have also been noticed for other similar Type of events 
\citep[e.g.][]{2006Natur.440..505L,2006A&A...454..827P}.
Here, we caution that there are large error bars associated with our polarization
measurements and also the amount of data is small. Additionally, the observed 
variations in the estimated values are at one sigma level and therefore, 
statistically insignificant.

If we examine the last three data points, it show a
constant level of polarization within one sigma limit with a mean value of $\sim$1.7 per cent.
This much level of polarization is comparable with some of the stripped envelope SNe such as
SN~1993J \citep{1996ApJ...459..307H,1997PASP..109..489T} and SN~2008D \citep{2010A&A...522A..14G}.
It is also notable that within the error limit, SN~2006ov and SN~2013ej show similar
level of polarization \citep[][]{2010ApJ...713.1363C} at later epochs. 
The increasing trend in the polarization light curve \citep[c.f.][]{2014MNRAS.442....2K}
of SN~2013ej has been seen after the end plateau phase, whereas, in case of SN~1987A 
and SN~1999em such features appeared before the end plateau phase ($\sim$40 d) and
continued up to late epochs \citep{1988MNRAS.234..937B,1991ApJS...77..405J,
2001ApJ...553..861L}.
The maximum level of polarization of SN~2008bk and SN~2004dj are almost similar.
Nevertheless, SN~2008bk shows an increasing trend during the plateau phase and then
in late epochs it retained nearly a constant polarization level. In contrary, SN~2004dj 
featured with different temporal behaviour. SN~2004dj exhibited almost negligible 
polarization in the similar interval, but once the inner core is revealed (end of 
plateau phase), the polarization abruptly increased and steadily decline towards
the nebular phase.
The increasing trend in polarization values of SN~2013ej has been noticed during 
the phase of transition from the plateau to nebular.

In continuation to above discussion, it should be emphasized that continuum polarization 
represents the underlying explosion geometry. Line polarization features on the other 
hand trace the asymmetries related to the relevant chemical elements in the SN ejecta
\citep{2003ApJ...593..788K,2008ARA&A..46..433W,2011MNRAS.415.3497D}. 
Local dust present around the SN could also result in optical scattering of the photons 
(i.e. ``light echo" mechanism) which may linearly polarize the reflected light.
\citet{1996ApJ...462L..27W} revisited the polarization properties of SN~1987A and favored 
this scenario \citep[see also][]{1996ApJ...459..307H,1997PASP..109..489T}. 
Restriction of data sample and single filter polarization analysis do not 
allow us to draw a definite conclusion on the geometry of SN~2013ej ejecta. 
Nonetheless, the contribution due to CSM interaction could be a plausible 
reason behind the polarization properties of this event.

Our analysis infer that possibly II-P SNe show diverse nature of ejecta. 
Post-plateau multi-epoch polarization data of similar events will be extremely useful 
which can further probe deeper into the ejecta and consequently may shed more light 
on the explosion geometry of these energetic events. In this context, along with existing 
large aperture observing facilities with spectropolarimetric capabilities (e.g. VLT, 
Subaru and Keck etc.), the proposed thirty meter telescope (TMT, \url{www.tmt.org/}) 
will provide extremely useful information about the geometry of CCSNe in the near future. 

%---------------------------------------------------------------------------------------
\section*{Acknowledgments}
This paper is dedicated to the memory of Dr. J. Gorosabel who had been our collaborator 
and contributed significantly towards the polarimetric study in astronomy. 
We are grateful to the observers Arti Joshi, B. J. Medhi and Ram Kesh Yadav at the 
Aryabhatta Research Institute of observational sciencES (ARIES) for their valuable 
time and support for the observations of this event. 
We acknowledge G. C. Anupama for careful reading of the manuscript and useful 
discussions on various polarimetric aspects.
BK acknowledges the financial support funded by the Canadian grant for the International Liquid
Mirror Telescope (ILMT) project during his stay at ARIES, India. 
BK also thanks Ravi Joshi for valuable discussions.
CE acknowledges the financial support from the grants MOST 103-2112-M-008-024-MY3 (at NCU, Taiwan) 
and MOST 102-2119-M-007-004-MY3 (at NTHU, Taiwan) funded by the Ministry of Science and Technology
of Taiwan.
This research has made use of the SIMBAD database, operated at CDS, Strasbourg, France. 

\label{lastpage}
\bibliography{sn13ej}

\begin{thebibliography}{122}
\expandafter\ifx\csname natexlab\endcsname\relax\def\natexlab#1{#1}\fi

\bibitem[{Anderson} et~al.(2014){Anderson}, {Gonz{\'a}lez-Gait{\'a}n}, {Hamuy}
  et~al.]{2014ApJ...786...67A}
{Anderson} J.~P., {Gonz{\'a}lez-Gait{\'a}n} S., {Hamuy} M., et~al., 2014, \apj,
  786, 67

\bibitem[{Anderson} et~al.(2011){Anderson}, {Habergham} \&
  {James}]{2011MNRAS.416..567A}
{Anderson} J.~P., {Habergham} S.~M., {James} P.~A., 2011, \mnras, 416, 567

\bibitem[{Anderson} \& {James}(2009)]{2009MNRAS.399..559A}
{Anderson} J.~P., {James} P.~A., 2009, \mnras, 399, 559

\bibitem[{Anderson} \& {Soto}(2013)]{2013A&A...550A..69A}
{Anderson} J.~P., {Soto} M., 2013, \aap, 550, A69

\bibitem[{Arnett}(1996)]{1996snih.book.....A}
{Arnett} D., 1996, {Supernovae and Nucleosynthesis: An Investigation of the
  History of Matter from the Big Bang to the Present (Princeton University
  Press)}

\bibitem[{Auld} et~al.(2006){Auld}, {Minchin}, {Davies}
  et~al.]{2006MNRAS.371.1617A}
{Auld} R., {Minchin} R.~F., {Davies} J.~I., et~al., 2006, \mnras, 371, 1617

\bibitem[{Barbon} et~al.(1979){Barbon}, {Ciatti} \&
  {Rosino}]{1979A&A....72..287B}
{Barbon} R., {Ciatti} F., {Rosino} L., 1979, \aap, 72, 287

\bibitem[{Barrett}(1988)]{1988MNRAS.234..937B}
{Barrett} P., 1988, \mnras, 234, 937

\bibitem[{Berger} et~al.(2002){Berger}, {Kulkarni} \&
  {Chevalier}]{2002ApJ...577L...5B}
{Berger} E., {Kulkarni} S.~R., {Chevalier} R.~A., 2002, \apjl, 577, L5

\bibitem[{Bose} \& {Kumar}(2014)]{2014ApJ...782...98B}
{Bose} S., {Kumar} B., 2014, \apj, 782, 98

\bibitem[{Bose} et~al.(2015){Bose}, {Sutaria}, {Kumar}
  et~al.]{2015ApJ...806..160B}
{Bose} S., {Sutaria} F., {Kumar} B., et~al., 2015, \apj, 806, 160

\bibitem[{Cardelli} et~al.(1989){Cardelli}, {Clayton} \&
  {Mathis}]{1989ApJ...345..245C}
{Cardelli} J.~A., {Clayton} G.~C., {Mathis} J.~S., 1989, \apj, 345, 245

\bibitem[{Chornock} et~al.(2010){Chornock}, {Filippenko}, {Li} \&
  {Silverman}]{2010ApJ...713.1363C}
{Chornock} R., {Filippenko} A.~V., {Li} W., {Silverman} J.~M., 2010, \apj, 713,
  1363

\bibitem[{Chugai}(2006)]{2006AstL...32..739C}
{Chugai} N.~N., 2006, Astronomy Letters, 32, 739

\bibitem[{Cornett} et~al.(1994){Cornett}, {O'Connell}, {Greason}
  et~al.]{1994ApJ...426..553C}
{Cornett} R.~H., {O'Connell} R.~W., {Greason} M.~R., et~al., 1994, \apj, 426,
  553

\bibitem[{Dessart} \& {Hillier}(2005)]{2005A&A...439..671D}
{Dessart} L., {Hillier} D.~J., 2005, \aap, 439, 671

\bibitem[{Dessart} \& {Hillier}(2011{\natexlab{a}})]{2011MNRAS.410.1739D}
{Dessart} L., {Hillier} D.~J., 2011{\natexlab{a}}, \mnras, 410, 1739

\bibitem[{Dessart} \& {Hillier}(2011{\natexlab{b}})]{2011MNRAS.415.3497D}
{Dessart} L., {Hillier} D.~J., 2011{\natexlab{b}}, \mnras, 415, 3497

\bibitem[{Eastman} et~al.(1996){Eastman}, {Schmidt} \&
  {Kirshner}]{1996ApJ...466..911E}
{Eastman} R.~G., {Schmidt} B.~P., {Kirshner} R., 1996, \apj, 466, 911

\bibitem[{Eldridge} et~al.(2013){Eldridge}, {Fraser}, {Smartt}, {Maund} \&
  {Crockett}]{2013MNRAS.436..774E}
{Eldridge} J.~J., {Fraser} M., {Smartt} S.~J., {Maund} J.~R., {Crockett} R.~M.,
  2013, \mnras, 436, 774

\bibitem[{Elmegreen} et~al.(2006){Elmegreen}, {Elmegreen}, {Chandar},
  {Whitmore} \& {Regan}]{2006ApJ...644..879E}
{Elmegreen} B.~G., {Elmegreen} D.~M., {Chandar} R., {Whitmore} B., {Regan} M.,
  2006, \apj, 644, 879

\bibitem[{Elmhamdi} et~al.(2003){Elmhamdi}, {Danziger}, {Chugai}
  et~al.]{2003MNRAS.338..939E}
{Elmhamdi} A., {Danziger} I.~J., {Chugai} N., et~al., 2003, \mnras, 338, 939

\bibitem[{Eswaraiah} et~al.(2013){Eswaraiah}, {Maheswar}, {Pandey}, {Jose},
  {Ramaprakash} \& {Bhatt}]{2013A&A...556A..65E}
{Eswaraiah} C., {Maheswar} G., {Pandey} A.~K., {Jose} J., {Ramaprakash} A.~N.,
  {Bhatt} H.~C., 2013, \aap, 556, A65

\bibitem[{Eswaraiah} et~al.(2011){Eswaraiah}, {Pandey}, {Maheswar}
  et~al.]{2011MNRAS.411.1418E}
{Eswaraiah} C., {Pandey} A.~K., {Maheswar} G., et~al., 2011, \mnras, 411, 1418

\bibitem[{Evans} \& {McNaught}(2003)]{2003IAUC.8150....2E}
{Evans} R., {McNaught} R.~H., 2003, \iaucirc, 8150, 2

\bibitem[{Falk} \& {Arnett}(1977)]{1977ApJS...33..515F}
{Falk} S.~W., {Arnett} W.~D., 1977, \apjs, 33, 515

\bibitem[{Filippenko}(1997)]{1997ARA&A..35..309F}
{Filippenko} A.~V., 1997, \araa, 35, 309

\bibitem[{Filippenko} \& {Chornock}(2002)]{2002IAUC.7825....1F}
{Filippenko} A.~V., {Chornock} R., 2002, \iaucirc, 7825, 1

\bibitem[{Foley} et~al.(2003){Foley}, {Papenkova}, {Swift}
  et~al.]{2003PASP..115.1220F}
{Foley} R.~J., {Papenkova} M.~S., {Swift} B.~J., et~al., 2003, \pasp, 115, 1220

\bibitem[{Fraser} et~al.(2014){Fraser}, {Maund}, {Smartt}
  et~al.]{2014MNRAS.439L..56F}
{Fraser} M., {Maund} J.~R., {Smartt} S.~J., et~al., 2014, \mnras, 439, L56

\bibitem[{Gal-Yam} et~al.(2002{\natexlab{a}}){Gal-Yam}, {Ofek} \&
  {Shemmer}]{2002MNRAS.332L..73G}
{Gal-Yam} A., {Ofek} E.~O., {Shemmer} O., 2002{\natexlab{a}}, \mnras, 332, L73

\bibitem[{Gal-Yam} et~al.(2002{\natexlab{b}}){Gal-Yam}, {Shemmer} \&
  {Dann}]{2002IAUC.7811....3G}
{Gal-Yam} A., {Shemmer} O., {Dann} J., 2002{\natexlab{b}}, \iaucirc, 7811, 3

\bibitem[{Gorosabel} et~al.(2010){Gorosabel}, {de Ugarte Postigo},
  {Castro-Tirado} et~al.]{2010A&A...522A..14G}
{Gorosabel} J., {de Ugarte Postigo} A., {Castro-Tirado} A.~J., et~al., 2010,
  \aap, 522, A14

\bibitem[{Grassberg} et~al.(1971){Grassberg}, {Imshennik} \&
  {Nadyozhin}]{1971Ap&SS..10...28G}
{Grassberg} E.~K., {Imshennik} V.~S., {Nadyozhin} D.~K., 1971, \apss, 10, 28

\bibitem[{Heger} et~al.(2003){Heger}, {Fryer}, {Woosley}, {Langer} \&
  {Hartmann}]{2003ApJ...591..288H}
{Heger} A., {Fryer} C.~L., {Woosley} S.~E., {Langer} N., {Hartmann} D.~H.,
  2003, \apj, 591, 288

\bibitem[{Heiles}(2000)]{2000AJ....119..923H}
{Heiles} C., 2000, \aj, 119, 923

\bibitem[{Hendry} et~al.(2005){Hendry}, {Smartt}, {Maund}
  et~al.]{2005MNRAS.359..906H}
{Hendry} M.~A., {Smartt} S.~J., {Maund} J.~R., et~al., 2005, \mnras, 359, 906

\bibitem[{Hoeflich}(1995)]{1995ApJ...440..821H}
{Hoeflich} P., 1995, \apj, 440, 821

\bibitem[{Hoeflich} et~al.(1996){Hoeflich}, {Wheeler}, {Hines} \&
  {Trammell}]{1996ApJ...459..307H}
{Hoeflich} P., {Wheeler} J.~C., {Hines} D.~C., {Trammell} S.~R., 1996, \apj,
  459, 307

\bibitem[{Hoflich}(1991)]{1991A&A...246..481H}
{Hoflich} P., 1991, \aap, 246, 481

\bibitem[{H{\o}g} et~al.(2000){H{\o}g}, {Fabricius}, {Makarov}
  et~al.]{2000A&A...355L..27H}
{H{\o}g} E., {Fabricius} C., {Makarov} V.~V., et~al., 2000, \aap, 355, L27

\bibitem[{Hsu} \& {Breger}(1982)]{1982ApJ...262..732H}
{Hsu} J.-C., {Breger} M., 1982, \apj, 262, 732

\bibitem[{Huang} et~al.(2015){Huang}, {Wang}, {Zhang}
  et~al.]{2015ApJ...807...59H}
{Huang} F., {Wang} X., {Zhang} J., et~al., 2015, \apj, 807, 59

\bibitem[{Iwamoto} et~al.(1998){Iwamoto}, {Mazzali}, {Nomoto}
  et~al.]{1998Natur.395..672I}
{Iwamoto} K., {Mazzali} P.~A., {Nomoto} K., et~al., 1998, \nat, 395, 672

\bibitem[{Iwamoto} et~al.(2000){Iwamoto}, {Nakamura}, {Nomoto}
  et~al.]{2000ApJ...534..660I}
{Iwamoto} K., {Nakamura} T., {Nomoto} K., et~al., 2000, \apj, 534, 660

\bibitem[{Jang} \& {Lee}(2014)]{2014ApJ...792...52J}
{Jang} I.~S., {Lee} M.~G., 2014, \apj, 792, 52

\bibitem[{Jeffery}(1991{\natexlab{a}})]{1991ApJ...375..264J}
{Jeffery} D.~J., 1991{\natexlab{a}}, \apj, 375, 264

\bibitem[{Jeffery}(1991{\natexlab{b}})]{1991ApJS...77..405J}
{Jeffery} D.~J., 1991{\natexlab{b}}, \apjs, 77, 405

\bibitem[{Jones} et~al.(2009){Jones}, {Hamuy}, {Lira}
  et~al.]{2009ApJ...696.1176J}
{Jones} M.~I., {Hamuy} M., {Lira} P., et~al., 2009, \apj, 696, 1176

\bibitem[{Kasen} et~al.(2003){Kasen}, {Nugent}, {Wang}
  et~al.]{2003ApJ...593..788K}
{Kasen} D., {Nugent} P., {Wang} L., et~al., 2003, \apj, 593, 788

\bibitem[{Kasen} et~al.(2006){Kasen}, {Thomas} \&
  {Nugent}]{2006ApJ...651..366K}
{Kasen} D., {Thomas} R.~C., {Nugent} P., 2006, \apj, 651, 366

\bibitem[{Kasen} \& {Woosley}(2009)]{2009ApJ...703.2205K}
{Kasen} D., {Woosley} S.~E., 2009, \apj, 703, 2205

\bibitem[{Kawabata} et~al.(2002){Kawabata}, {Jeffery}, {Iye}
  et~al.]{2002ApJ...580L..39K}
{Kawabata} K.~S., {Jeffery} D.~J., {Iye} M., et~al., 2002, \apjl, 580, L39

\bibitem[{Kilgard} et~al.(2005){Kilgard}, {Cowan}, {Garcia}
  et~al.]{2005ApJS..159..214K}
{Kilgard} R.~E., {Cowan} J.~J., {Garcia} M.~R., et~al., 2005, \apjs, 159, 214

\bibitem[{Kim} et~al.(2013){Kim}, {Zheng}, {Li} et~al.]{2013CBET.3606....1K}
{Kim} M., {Zheng} W., {Li} W., et~al., 2013, Central Bureau Electronic
  Telegrams, 3606, 1

\bibitem[{Kinugasa} et~al.(2002{\natexlab{a}}){Kinugasa}, {Kawakita}, {Ayani}
  et~al.]{2002ApJ...577L..97K}
{Kinugasa} K., {Kawakita} H., {Ayani} K., et~al., 2002{\natexlab{a}}, \apjl,
  577, L97

\bibitem[{Kinugasa} et~al.(2002{\natexlab{b}}){Kinugasa}, {Kawakita}, {Ayani},
  {Kawabata} \& {Yamaoka}]{2002IAUC.7811....1K}
{Kinugasa} K., {Kawakita} H., {Ayani} K., {Kawabata} T., {Yamaoka} H.,
  2002{\natexlab{b}}, \iaucirc, 7811, 1

\bibitem[{Kirshner} \& {Kwan}(1974)]{1974ApJ...193...27K}
{Kirshner} R.~P., {Kwan} J., 1974, \apj, 193, 27

\bibitem[{Krauss} et~al.(2005){Krauss}, {Kilgard}, {Garcia}, {Roberts} \&
  {Prestwich}]{2005ApJ...630..228K}
{Krauss} M.~I., {Kilgard} R.~E., {Garcia} M.~R., {Roberts} T.~P., {Prestwich}
  A.~H., 2005, \apj, 630, 228

\bibitem[{Kumar} et~al.(2014){Kumar}, {Pandey}, {Eswaraiah} \&
  {Gorosabel}]{2014MNRAS.442....2K}
{Kumar} B., {Pandey} S.~B., {Eswaraiah} C., {Gorosabel} J., 2014, \mnras, 442,
  2

\bibitem[{Leli{\`e}vre} \& {Roy}(2000)]{2000AJ....120.1306L}
{Leli{\`e}vre} M., {Roy} J.-R., 2000, \aj, 120, 1306

\bibitem[{Leonard} et~al.(2012){Leonard}, {Dessart}, {Hillier} \&
  {Pignata}]{2012AIPC.1429..204L}
{Leonard} D.~C., {Dessart} L., {Hillier} D.~J., {Pignata} G., 2012, in {
  American Institute of Physics Conference Series\/}, edited by J.~L.
  {Hoffman}, J.~{Bjorkman}, B.~{Whitney}, vol. 1429 of { American Institute of
  Physics Conference Series\/},  204--207

\bibitem[{Leonard} \& {Filippenko}(2001)]{2001PASP..113..920L}
{Leonard} D.~C., {Filippenko} A.~V., 2001, \pasp, 113, 920

\bibitem[{Leonard} \& {Filippenko}(2005)]{2005ASPC..342..330L}
{Leonard} D.~C., {Filippenko} A.~V., 2005, in { 1604-2004: Supernovae as
  Cosmological Lighthouses\/}, edited by M.~{Turatto}, S.~{Benetti},
  L.~{Zampieri}, W.~{Shea}, vol. 342 of { Astronomical Society of the Pacific
  Conference Series\/},  330

\bibitem[{Leonard} et~al.(2001){Leonard}, {Filippenko}, {Ardila} \&
  {Brotherton}]{2001ApJ...553..861L}
{Leonard} D.~C., {Filippenko} A.~V., {Ardila} D.~R., {Brotherton} M.~S., 2001,
  \apj, 553, 861

\bibitem[{Leonard} et~al.(2002{\natexlab{a}}){Leonard}, {Filippenko},
  {Chornock} \& {Foley}]{2002PASP..114.1333L}
{Leonard} D.~C., {Filippenko} A.~V., {Chornock} R., {Foley} R.~J.,
  2002{\natexlab{a}}, \pasp, 114, 1333

\bibitem[{Leonard} et~al.(2002{\natexlab{b}}){Leonard}, {Filippenko},
  {Chornock} \& {Li}]{2002AJ....124.2506L}
{Leonard} D.~C., {Filippenko} A.~V., {Chornock} R., {Li} W.,
  2002{\natexlab{b}}, \aj, 124, 2506

\bibitem[{Leonard} et~al.(2006){Leonard}, {Filippenko}, {Ganeshalingam}
  et~al.]{2006Natur.440..505L}
{Leonard} D.~C., {Filippenko} A.~V., {Ganeshalingam} M., et~al., 2006, \nat,
  440, 505

\bibitem[{Leonard} et~al.(2002{\natexlab{c}}){Leonard}, {Filippenko}, {Li}
  et~al.]{2002AJ....124.2490L}
{Leonard} D.~C., {Filippenko} A.~V., {Li} W., et~al., 2002{\natexlab{c}}, \aj,
  124, 2490

\bibitem[{Leonard} et~al.(2013){Leonard}, {Pignata}, {Dessart}
  et~al.]{2013ATel.5275....1L}
{Leonard} P.~b.~D.~C., {Pignata} G., {Dessart} L., et~al., 2013, The
  Astronomer's Telegram, 5275, 1

\bibitem[{Li} et~al.(2000){Li}, {Filippenko}, {Treffers}
  et~al.]{2000AIPC..522..103L}
{Li} W.~D., {Filippenko} A.~V., {Treffers} R.~R., et~al., 2000, in { American
  Institute of Physics Conference Series\/}, edited by S.~S. {Holt}, W.~W.
  {Zhang}, vol. 522 of { American Institute of Physics Conference Series\/},
  103--106

\bibitem[{Maeda} et~al.(2006){Maeda}, {Mazzali} \&
  {Nomoto}]{2006ApJ...645.1331M}
{Maeda} K., {Mazzali} P.~A., {Nomoto} K., 2006, \apj, 645, 1331

\bibitem[{Mazzali} et~al.(2002){Mazzali}, {Deng}, {Maeda}
  et~al.]{2002ApJ...572L..61M}
{Mazzali} P.~A., {Deng} J., {Maeda} K., et~al., 2002, \apjl, 572, L61

\bibitem[{Mazzali} et~al.(2006){Mazzali}, {Deng}, {Pian}
  et~al.]{2006ApJ...645.1323M}
{Mazzali} P.~A., {Deng} J., {Pian} E., et~al., 2006, \apj, 645, 1323

\bibitem[{Mazzali} et~al.(2003){Mazzali}, {Deng}, {Tominaga}
  et~al.]{2003ApJ...599L..95M}
{Mazzali} P.~A., {Deng} J., {Tominaga} N., et~al., 2003, \apjl, 599, L95

\bibitem[{Meikle} et~al.(2002){Meikle}, {Lucy}, {Smartt}, {Leibundgut},
  {Lundqvist} \& {Ostensen}]{2002IAUC.7811....2M}
{Meikle} P., {Lucy} L., {Smartt} S., {Leibundgut} B., {Lundqvist} P.,
  {Ostensen} R., 2002, \iaucirc, 7811, 2

\bibitem[{Nakano} et~al.(2002){Nakano}, {Hirose}, {Kushida}, {Kushida} \&
  {Li}]{2002IAUC.7810....1N}
{Nakano} S., {Hirose} Y., {Kushida} R., {Kushida} Y., {Li} W., 2002, \iaucirc,
  7810, 1

\bibitem[{Pandey} et~al.(2009){Pandey}, {Medhi}, {Sagar} \&
  {Pandey}]{2009MNRAS.396.1004P}
{Pandey} J.~C., {Medhi} B.~J., {Sagar} R., {Pandey} A.~K., 2009, \mnras, 396,
  1004

\bibitem[{Pandey} et~al.(2003{\natexlab{a}}){Pandey}, {Anupama}, {Sagar},
  {Bhattacharya}, {Sahu} \& {Pandey}]{2003MNRAS.340..375P}
{Pandey} S.~B., {Anupama} G.~C., {Sagar} R., {Bhattacharya} D., {Sahu} D.~K.,
  {Pandey} J.~C., 2003{\natexlab{a}}, \mnras, 340, 375

\bibitem[{Pandey} et~al.(2003{\natexlab{b}}){Pandey}, {Sahu}, {Anupama},
  {Bhattacharya} \& {Sagar}]{2003BASI...31..351P}
{Pandey} S.~B., {Sahu} D.~K., {Anupama} G.~C., {Bhattacharya} D., {Sagar} R.,
  2003{\natexlab{b}}, Bulletin of the Astronomical Society of India, 31, 351

\bibitem[{Patat} \& {Romaniello}(2006)]{2006PASP..118..146P}
{Patat} F., {Romaniello} M., 2006, \pasp, 118, 146

\bibitem[{Patat} et~al.(2011){Patat}, {Taubenberger}, {Benetti}, {Pastorello}
  \& {Harutyunyan}]{2011A&A...527L...6P}
{Patat} F., {Taubenberger} S., {Benetti} S., {Pastorello} A., {Harutyunyan} A.,
  2011, \aap, 527, L6

\bibitem[{Pejcha} \& {Prieto}(2015)]{2015ApJ...806..225P}
{Pejcha} O., {Prieto} J.~L., 2015, \apj, 806, 225

\bibitem[{Pereyra} et~al.(2006){Pereyra}, {Magalh{\~a}es}, {Rodrigues}
  et~al.]{2006A&A...454..827P}
{Pereyra} A., {Magalh{\~a}es} A.~M., {Rodrigues} C.~V., et~al., 2006, \aap,
  454, 827

\bibitem[{Phillips} et~al.(1990){Phillips}, {Hamuy}, {Heathcote}, {Suntzeff} \&
  {Kirhakos}]{1990AJ.....99.1133P}
{Phillips} M.~M., {Hamuy} M., {Heathcote} S.~R., {Suntzeff} N.~B., {Kirhakos}
  S., 1990, \aj, 99, 1133

\bibitem[{Pignata}(2013)]{2013msao.confE.176P}
{Pignata} G., 2013, in { Massive Stars: From alpha to Omega\/},  176

\bibitem[{Popov}(1993)]{1993ApJ...414..712P}
{Popov} D.~V., 1993, \apj, 414, 712

\bibitem[{Rautela} et~al.(2004){Rautela}, {Joshi} \&
  {Pandey}]{2004BASI...32..159R}
{Rautela} B.~S., {Joshi} G.~C., {Pandey} J.~C., 2004, Bulletin of the
  Astronomical Society of India, 32, 159

\bibitem[{Richmond}(2014)]{2014JAVSO..42..333R}
{Richmond} M.~W., 2014, Journal of the American Association of Variable Star
  Observers (JAAVSO), 42, 333

\bibitem[{Schlafly} \& {Finkbeiner}(2011)]{2011ApJ...737..103S}
{Schlafly} E.~F., {Finkbeiner} D.~P., 2011, \apj, 737, 103

\bibitem[{Schmidt} et~al.(1992){Schmidt}, {Elston} \&
  {Lupie}]{1992AJ....104.1563S}
{Schmidt} G.~D., {Elston} R., {Lupie} O.~L., 1992, \aj, 104, 1563

\bibitem[{Serkowski}(1958)]{1958AcA.....8..135S}
{Serkowski} K., 1958, \actaa, 8, 135

\bibitem[{Serkowski}(1970)]{1970ApJ...160.1083S}
{Serkowski} K., 1970, \apj, 160, 1083

\bibitem[{Serkowski} et~al.(1975){Serkowski}, {Mathewson} \&
  {Ford}]{1975ApJ...196..261S}
{Serkowski} K., {Mathewson} D.~S., {Ford} V.~L., 1975, \apj, 196, 261

\bibitem[{Shapiro} \& {Sutherland}(1982)]{1982ApJ...263..902S}
{Shapiro} P.~R., {Sutherland} P.~G., 1982, \apj, 263, 902

\bibitem[{Skrutskie} et~al.(2006){Skrutskie}, {Cutri}, {Stiening}
  et~al.]{2006AJ....131.1163S}
{Skrutskie} M.~F., {Cutri} R.~M., {Stiening} R., et~al., 2006, \aj, 131, 1163

\bibitem[{Smartt} et~al.(2009){Smartt}, {Eldridge}, {Crockett} \&
  {Maund}]{2009MNRAS.395.1409S}
{Smartt} S.~J., {Eldridge} J.~J., {Crockett} R.~M., {Maund} J.~R., 2009,
  \mnras, 395, 1409

\bibitem[{Smith} et~al.(2011){Smith}, {Li}, {Filippenko} \&
  {Chornock}]{2011MNRAS.412.1522S}
{Smith} N., {Li} W., {Filippenko} A.~V., {Chornock} R., 2011, \mnras, 412, 1522

\bibitem[{Soam} et~al.(2015){Soam}, {Maheswar}, {Lee}
  et~al.]{2015A&A...573A..34S}
{Soam} A., {Maheswar} G., {Lee} C.~W., et~al., 2015, \aap, 573, A34

\bibitem[{Soderberg} et~al.(2006){Soderberg}, {Nakar}, {Berger} \&
  {Kulkarni}]{2006ApJ...638..930S}
{Soderberg} A.~M., {Nakar} E., {Berger} E., {Kulkarni} S.~R., 2006, \apj, 638,
  930

\bibitem[{Sonba{\c s}} et~al.(2010){Sonba{\c s}}, {Aky{\"u}z}, {Balman} \&
  {{\"O}zel}]{2010A&A...517A..91S}
{Sonba{\c s}} E., {Aky{\"u}z} A., {Balman} {\c S}., {{\"O}zel} M.~E., 2010,
  \aap, 517, A91

\bibitem[{Soria} \& {Kong}(2002)]{2002ApJ...572L..33S}
{Soria} R., {Kong} A.~K.~H., 2002, \apjl, 572, L33

\bibitem[{Soria} et~al.(2004){Soria}, {Pian} \& {Mazzali}]{2004A&A...413..107S}
{Soria} R., {Pian} E., {Mazzali} P.~A., 2004, \aap, 413, 107

\bibitem[{Spiro} et~al.(2014){Spiro}, {Pastorello}, {Pumo}
  et~al.]{2014MNRAS.439.2873S}
{Spiro} S., {Pastorello} A., {Pumo} M.~L., et~al., 2014, \mnras, 439, 2873

\bibitem[{Sutaria} et~al.(2003){Sutaria}, {Chandra}, {Bhatnagar} \&
  {Ray}]{2003A&A...397.1011S}
{Sutaria} F.~K., {Chandra} P., {Bhatnagar} S., {Ray} A., 2003, \aap, 397, 1011

\bibitem[{Tran}(1995)]{1995ApJ...440..565T}
{Tran} H.~D., 1995, \apj, 440, 565

\bibitem[{Tran} et~al.(1997){Tran}, {Filippenko}, {Schmidt}, {Bjorkman},
  {Jannuzi} \& {Smith}]{1997PASP..109..489T}
{Tran} H.~D., {Filippenko} A.~V., {Schmidt} G.~D., {Bjorkman} K.~S., {Jannuzi}
  B.~T., {Smith} P.~S., 1997, \pasp, 109, 489

\bibitem[{Utrobin}(2007)]{2007A&A...461..233U}
{Utrobin} V.~P., 2007, \aap, 461, 233

\bibitem[{Utrobin} \& {Chugai}(2015)]{2015A&A...575A.100U}
{Utrobin} V.~P., {Chugai} N.~N., 2015, \aap, 575, A100

\bibitem[{Valenti} et~al.(2013){Valenti}, {Sand}, {Howell}
  et~al.]{2013CBET.3609....1V}
{Valenti} S., {Sand} D., {Howell} D.~A., et~al., 2013, Central Bureau
  Electronic Telegrams, 3609, 1

\bibitem[{Valenti} et~al.(2014){Valenti}, {Sand}, {Pastorello}
  et~al.]{2014MNRAS.438L.101V}
{Valenti} S., {Sand} D., {Pastorello} A., et~al., 2014, \mnras, 438, L101

\bibitem[{van Leeuwen}(2007)]{2007A&A...474..653V}
{van Leeuwen} F., 2007, \aap, 474, 653

\bibitem[{Vink{\'o}} et~al.(2004){Vink{\'o}}, {Blake}, {S{\'a}rneczky}
  et~al.]{2004A&A...427..453V}
{Vink{\'o}} J., {Blake} R.~M., {S{\'a}rneczky} K., et~al., 2004, \aap, 427, 453

\bibitem[{Vink{\'o}} et~al.(2009){Vink{\'o}}, {S{\'a}rneczky}, {Balog}
  et~al.]{2009ApJ...695..619V}
{Vink{\'o}} J., {S{\'a}rneczky} K., {Balog} Z., et~al., 2009, \apj, 695, 619

\bibitem[{Wang} et~al.(2003){Wang}, {Baade}, {H{\"o}flich} \&
  {Wheeler}]{2003ApJ...592..457W}
{Wang} L., {Baade} D., {H{\"o}flich} P., {Wheeler} J.~C., 2003, \apj, 592, 457

\bibitem[{Wang} \& {Wheeler}(1996)]{1996ApJ...462L..27W}
{Wang} L., {Wheeler} J.~C., 1996, \apjl, 462, L27

\bibitem[{Wang} \& {Wheeler}(2008)]{2008ARA&A..46..433W}
{Wang} L., {Wheeler} J.~C., 2008, \araa, 46, 433

\bibitem[{Wardle} \& {Kronberg}(1974)]{1974ApJ...194..249W}
{Wardle} J.~F.~C., {Kronberg} P.~P., 1974, \apj, 194, 249

\bibitem[{Wheeler}(2000)]{2000ccsg.book.....W}
{Wheeler} J.~C., 2000, {Cosmic catastrophes: supernovae, gamma-ray bursts, and
  adventures in hyperspace, Cambridge University Press}

\bibitem[{Wheeler} \& {Filippenko}(1996)]{1996ssr..conf..241W}
{Wheeler} J.~C., {Filippenko} A.~V., 1996, in { IAU Colloq. 145: Supernovae and
  Supernova Remnants\/}, edited by T.~S. {Kuhn},  241

\bibitem[{Whittet}(2003)]{2003dge..conf.....W}
{Whittet} D.~C.~B., 2003, {Dust in the galactic environment (Bristol: IOP)}

\bibitem[{Zhang} et~al.(2006){Zhang}, {Wang}, {Li} et~al.]{2006AJ....131.2245Z}
{Zhang} T., {Wang} X., {Li} W., et~al., 2006, \aj, 131, 2245

\end{thebibliography}

\appendix 

\section{Estimation of polarization contamination} \label{apn_a}

In this section we briefly describe about our procedures to estimate the possible 
contaminations introduced due to diffuse background, during different phases of 
moon and/or seeing conditions.
These effects could contribute to the polarization values in addition to those 
contributed by the ISP.  

\begin{figure}
\begin{centering}
\includegraphics[scale=0.11]{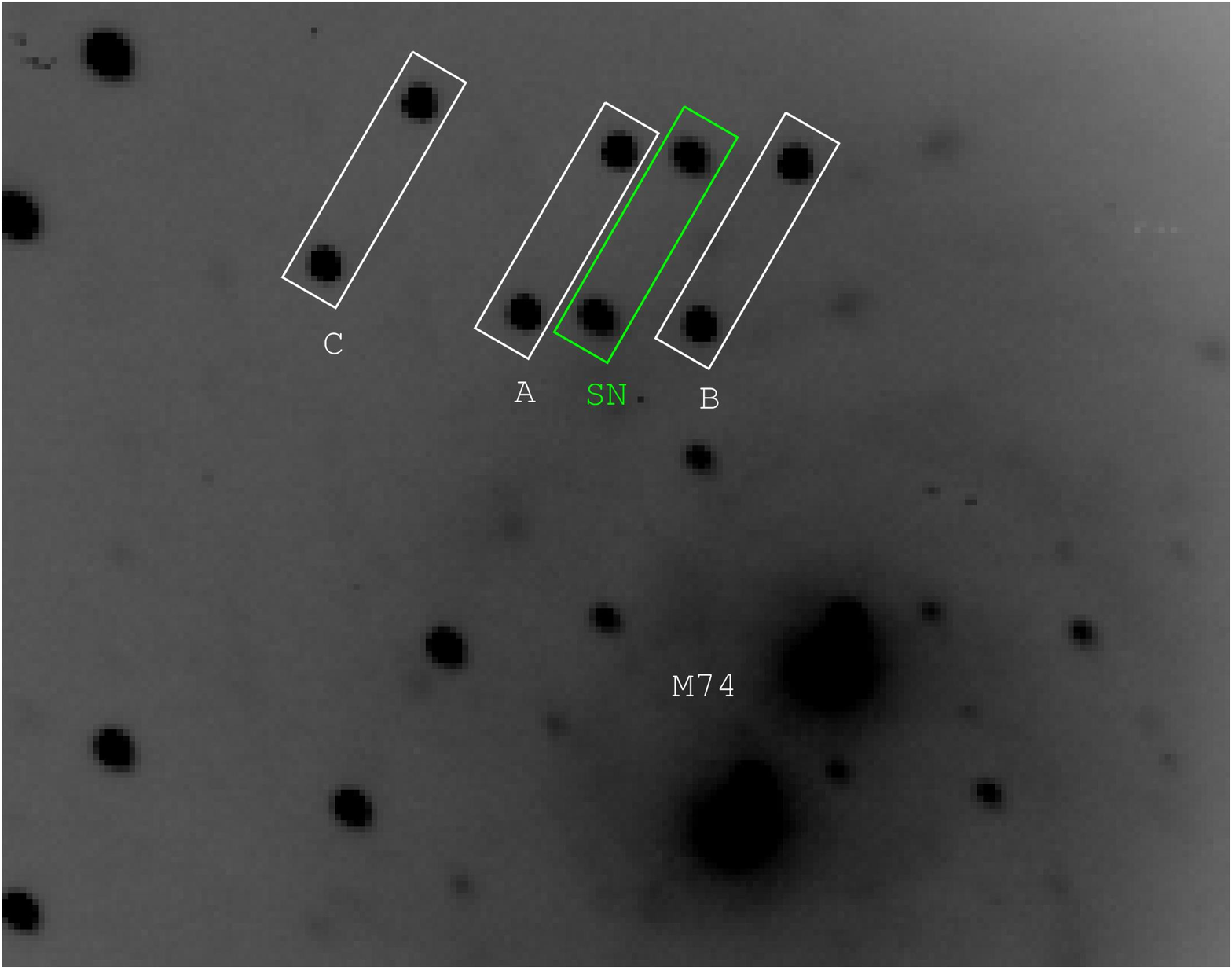}
\caption{Co-added image of SN~2013ej and field star HD~8919. 
The SN position is marked with ``SN" (green rectangle) and co-added positions 
of field star are labelled with ``A, B and C" (white rectangles). Host galaxy 
is also indicated. Two images of each object represent ordinary and extra-ordinary image.}
\label{backg}
\end{centering}
\end{figure}

Initially, the SN images acquired at each epoch were aligned using {\small {\it imalign}} task 
in {\small IRAF} followed by preprocessing (see Section~\ref{obs}).
We visually selected three locations A, B and C over one of these aligned images (see Fig.~\ref{backg}). 
These locations were chosen in such a way that A, B and C are lying nearest to farthest angular 
distance from the SN. Out of them B and C belong to most and least nebulous region, respectively. 
Generally, well isolated field stars may not have diffuse background as they are well separated 
from the nebulous regions (e.g. H\,{\small II region}). 
Estimation of polarization parameters at various positions (in this case A, B and C) of such star
will indicate variation in these values and finally provide possible contribution of nebulous 
background.   

\begin{figure}
\begin{centering}
\includegraphics[scale=0.4]{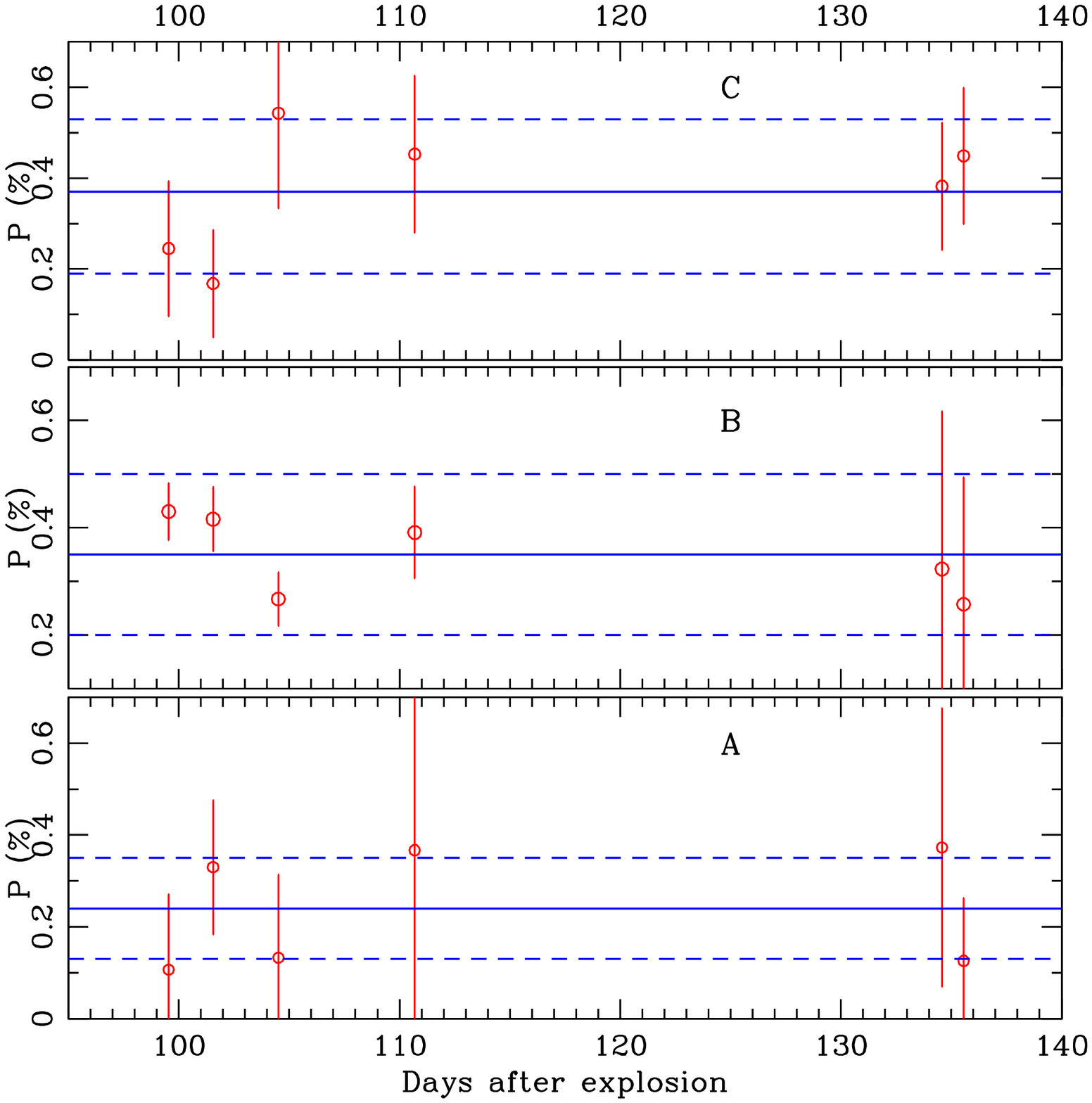}
\caption{Estimated polarization values of field star at three different positions after 
co-addition with the SN and HD~8919 images. Panels A, B and C respectively, indicate 
variation in the degree of polarization at various epoch of observations for three marked 
positions in Fig.~\ref{backg}. 
The blue coloured continuous and dashed lines (in each panel) respectively, indicate the 
mean value of degree of polarization and one sigma uncertainties after co-addition of the 
images.  
}
\label{backg_plot}
\end{centering}
\end{figure}

Therefore, we focused to field star HD~8919 which was already observed by AIMPOL with 
$P=0.29 \pm 0.06$ (see Table~\ref{tab:field_stars}). Situated at a distance of 525 pc, 
HD~8919 is most distant and faintest \citep[$V \sim$10.07 mag,][]{2000A&A...355L..27H} 
object in our list of field stars (c.f. Table~\ref{tab:field_stars}). 
The corresponding four images (taken at each HWP positions) of HD~8919 and SN were co-added.
It was done by {\small {\it imcombine}} task in {\small IRAF}.  
Then, we estimated the polarization values of this star at every epoch by applying the same 
offset and aperture values by which SN polarization values were estimated. The results evaluated 
at each A, B and C positions are overplotted in panels A, B and C in Fig.~\ref{backg_plot}. 
We found a mean polarization value
of 0.24 $\pm$ 0.11, 0.35 $\pm$ 0.15 and 0.37 $\pm$ 0.16 for A, B and C positions, respectively.
The blue coloured continuous and dashed lines in each panel respectively, indicate the
mean degree of polarization and one sigma uncertainties.

We do not see significant variation in epoch to epoch values for three positions and also 
these are lying almost within the one sigma uncertainty limit. This suggests that resultant 
of nebulous background may not have significant effect on the final measurements of SN.
It should be noted that although reported method may not be very accurate but could be 
useful to verify the background contribution for the objects overimposed to the nebulous 
regions.  

\end{document}